\documentclass{aa}

\usepackage{graphicx}
\usepackage{txfonts}
\usepackage[colorlinks,citecolor=blue]{hyperref}
\hypersetup{
    colorlinks = true,
    linkcolor = blue,
    citecolor = blue,
    filecolor = blue,
    urlcolor = blue
    }

\usepackage{mathrsfs}
\usepackage{booktabs}
\usepackage{mathtools}
\usepackage{placeins}
\usepackage{siunitx}
\DeclareSIUnit{\dex}{dex}
\DeclareSIUnit{\Msun}{\textit{M}_\odot}
\DeclareSIUnit{\Rsun}{\textit{R}_\odot}
\DeclareSIUnit{\Lsun}{\textit{L}_\odot}
\DeclareSIUnit{\pc}{pc}
\DeclareSIUnit{\mas}{mas}
\DeclareSIUnit{\year}{yr}
\DeclareSIUnit{\mag}{mag}
\DeclareSIUnit{\au}{AU}
\usepackage{svg}
\usepackage{ulem}

\begin{document} 

   \title{Quantitative spectroscopy of multiple OB stars\\ I. The quadruple system HD~37061 at the centre of Messier~43}

   \titlerunning{The quadruple system HD~37061 at the centre of Messier~43}

   \author{P. Aschenbrenner\inst{1}
          \and
          N. Przybilla\inst{1}
          }

   \institute{Universit\"at Innsbruck, Institut f\"ur Astro- und Teilchenphysik, Technikerstr. 25/8, 6020 Innsbruck, Austria\\
              \email{patrick.aschenbrenner@student.uibk.ac.at; norbert.przybilla@uibk.ac.at}
             }

   \date{Received ; accepted }

  \abstract
   {The majority of massive stars are located in binary or multiple star systems. Compared to single stars, these objects pose additional challenges to quantitative analyses based on model atmospheres. In particular, little information is currently available on the chemical composition of such systems.}
   {The members of the quadruple star system HD~37061, which excites the \ion{H}{ii} region Messier~43 in Orion, are fully characterised. Accurate and precise abundances for all elements with lines traceable in the optical spectrum are derived for the first time.}
   {A hybrid non-local thermodynamic equilibrium (non-LTE) approach, using line-blanketed hydrostatic model atmospheres computed with the {\sc Atlas12} code in combination with non-LTE line-formation calculations with {\sc Detail} and {\sc Surface}, was employed. 
   A high-resolution composite spectrum was analysed for the atmospheric parameters and elemental abundances of the individual stars. Fundamental stellar parameters were derived based on stellar evolution tracks, and the interstellar reddening was characterised.} 
   {We determined the fundamental parameters and chemical abundances for three stars in the HD\,37061 system.
   The fourth and faintest star in the system shows no distinct spectral features, as a result of its fast rotation.
   However, this star has noticeable effects on the continuum.
   The derived element abundances and determined ages of the individual stars are consistent with each other, and the abundances coincide with the cosmic abundance standard.
   We find an excellent agreement between our spectroscopic distance and the \textit{Gaia} Data Release~3 parallax distance.}
   {}


   \maketitle

\section{Introduction}
The majority of massive stars, that is, stars with initial masses higher than $\sim$8\,$M_\odot$, are situated in binary or multiple star systems \citep[e.g.][]{Chinietal12,Sanaetal12}. Binary stars are the main sources of fundamental data on stellar masses,~$M$, and radii, $R$, and are therefore pivotal for the empirical calibration of stellar evolution models. For detached double-lined eclipsing binaries (DEBs) an accuracy of better than $\pm$3\% for both $M$ and $R$ can be achieved, as the results rely on only geometry and Newtonian mechanics \citep{Andersen91,Torresetal10}.

While a number of studies on fundamental stellar parameters are available (see the references in the Andersen and Torres et al. work, as well as more examples of recent work referred to in this paragraph), the knowledge of elemental surface abundances in binary and multiple OB star systems is sparse. This is despite their importance in the context of stellar evolution (tracing mixing processes that can bring nuclear-processed matter to the surface layers) and galactochemical evolution (tracing the chemical homogeneity of stellar environments on small scales and galactic abundance gradients on large scales). Elemental abundances have been routinely determined for (preferentially slowly rotating) single stars or single-lined binary (SB1) stars \citep[for more recent results, see e.g.][]{Hunteretal09,NiSi11,NiPr12,Martinsetal17,Carneiroetal19,Moreletal22,Blommeetal22,Aschenbrenneretal23}.
Yet, the composite spectra of double-lined systems (SB2) or even more stars pose additional challenges;
as a result, full quantitative spectral analyses of binary or multiple OB stars remain relatively rare. So far, model atmosphere analyses of binaries have been performed on the basis of disentangled spectra from a series of observations \citep{SiSt94,Hadrava95}, effectively reducing the problem to the standard task of analysing single-star spectra. This encompasses both non-eclipsing \citep[e.g.][]{Mahyetal10,Tkatchenkoetal16,Raucqetal18,Fabryetal21} and DEB systems \citep[e.g.][]{PaHe05,PaSo09,Pavlovskietal18,Pavlovskietal23,Johnstonetal19}. However, the number of chemical elements investigated so far has been restricted to a handful at most, concentrating in particular on C, N, and O to study the effects of mixing in the course of the evolution of the systems' stars \citep[see in particular][]{Pavlovskietal23}.

A different approach, which in principle can be applied to a single-epoch spectrum of a binary or multiple OB star system, was introduced by \citet{Irrgangetal14}, who studied ten metal species that produce spectral lines in the visual spectrum. Such a single-epoch spectral analysis approach was cross-matched with the analysis of disentangled time-series spectra for a triple star system and a binary with one pulsating component, both of which include one massive magnetic star \citep{Gonzalezetal17,Gonzalezetal19}. What is missing is a more systematic investigation of binary and multiple OB systems, to study to what degree the components share the same abundances (stemming from the same birth cloud) or to what degree chemical peculiarities may develop (e.g. in the presence of a magnetic field). 

With the present work, we begin this endeavour, taking advantage of further constraints that have become available over the years from high-angular-resolution observations that use adaptive optics and interferometric techniques as well as \textit{Gaia} data. Thanks to the many improvements to our analysis methodology that permit an increasingly close reproduction of the optical spectra of OB stars, we intend to provide abundances with uncertainties (1$\sigma$ standard deviations) smaller than 0.1\,dex so that robust conclusions can be drawn.

The object \object{HD 37061} (NU Ori) is the central ionising star of the compact and spherical \ion{H}{ii} region Messier~43 in Orion. The spectral classification ranges from O9\,V \citep{Bragancaetal12} to B1\,V \citep{Feigelsonetal02}.
We use the modern designation of individual stars by \cite{Shultzetal19} in in this paper.
A close spectroscopic binary was first detected by \cite{MorrellLevato91}. Based on radial velocity measurements \cite{Abtetal91} determined an orbital solution with a period of $\sim$$\SI{19}{\day}$ (primary designated HD~37061\,Aa, the secondary HD~37061\,Ab). Later, \cite{Preibischetal99} discovered a second companion at a distance of $\sim$$\SI{470}{mas}$, designated HD~37061\,B. Indications of the presence of a fourth star in the system were found by \cite{Grellmannetal13}, and this object was confirmed by \cite{GravityCol1aboration18} at a distance of $\sim$$\SI{8.6}{mas}$ and designated HD~37061\,C. 

Over the past few decades, HD~37061 has attracted interest mostly as a bright background source used to study the interstellar medium (ISM) along an interesting line of sight. Quantitative spectroscopy of HD~37061 has rarely been performed. \citet{Simon-Diazetal11} analysed an intermediate-resolution spectrum, investigating hydrogen and helium lines and accounting only for the primary star.
A detailed determination of the orbital elements and fundamental parameters of the three innermost stars was made by \cite{Shultzetal19}. They used radial velocity measurements from high-resolution spectra and find orbital periods of $\sim$$\SI{14}{\day}$ and $\sim$$\SI{476}{\day}$, respectively. The star HD~37061\,C is identified as the magnetic star of the system in that work. The chemical composition of the system's stars has not yet been determined.

The paper is structured as follows.
Observations and data reduction are presented in Sect. \ref{section:observations}.
A description of our models and applied methods is given in Sect. \ref{section:model_atmospheres}.
The individual steps of our quantitative analysis are summarised in Sect. \ref{section:spectral_analysis}, and the results, including for the first time a chemical analysis of the system, are discussed in Sect. \ref{section:results}.
Finally, we give a summary in Sect. \ref{section:summary}.
Appendix \ref{appendix:A} shows fits to the spectrum of HD~37061.

\section{Observations and data reduction}\label{section:observations}
The analysis is based on high-resolution spectra observed with the
Echelle Spectro-Polari\-metric Device
for the Observation of Stars \citep[{ESPaDOnS},][]{ManDon03} on the
3.6\,m Canada-France-Hawaii Telescope (CFHT) on Mauna Kea in Hawaii.
The spectra cover a wavelength range from 3700 to 10\,500\,{\AA} at a resolving 
power of $R$\,=\,$\lambda / \Delta \lambda$\,$\approx$\,68\,000. We used
pipeline-reduced spectra downloaded from the CFHT Science Archive at the Canadian Astronomy
Data Centre\footnote{\url{https://www.cadc-ccda.hia-iha.nrc-cnrc.gc.ca/en/cfht/}}.
The spectra were normalised by fitting a spline function through carefully
selected continuum points. We co-added three spectra observed within $2.5$ hours on March 8, 2007, to achieve a signal-to-noise ratio (S/N) of 800, measured at $\sim$5100\,{\AA}.
In addition, various (spectrophotometric) data were used. Low-dispersion, large-aperture 
spectra taken with the International Ultraviolet Explorer (IUE; see Table~\ref{table:IUE_data})
were downloaded from the Mikulski Archive for Space Telescopes (MAST\footnote{\url{https://archive.stsci.edu/iue/}}).
The short-wavelength (SW) data cover the range 
$\lambda\lambda$1150-1978\,{\AA} and the long-wavelength (LW) data $\lambda\lambda$1851-3347\,{\AA}.
The IUE spectrophotometric data were co-added to increase the S/N. We note that the radial velocity variations of the spectrum have no effect on the continuum shape, which is later employed  to investigate the spectral energy distribution (SED).

\begin{table}
\caption{IUE spectrophotometry used in the present work.}
\label{table:IUE_data}      
\centering  
    \resizebox{\linewidth}{!}{\small\begin{tabular}{llcccc} 
\hline\hline                
Object    & SW     & Date       & LW     & Date       \\ \hline
 HD 37061  & P05084 & 1979-04-29 & R05970 & 1979-10-29 \\
          & P06954 & 1979-10-22 & R10433 & 1981-04-24 \\
          & P13799 & 1981-04-24 & &  \\
 \hline
\end{tabular}}
\end{table}

\section{Model atmospheres and spectrum synthesis}\label{section:model_atmospheres}

\subsection{Models and programs}
{The analysis is based on a hybrid non-local thermodynamic equilibrium (non-LTE) approach. We employed 
non-LTE line-formation calculations with the updated and extended versions of the codes {\sc Detail} and {\sc Surface} \citep{giddings81,BuGi85} based on LTE model atmospheres, computed with the {\sc Atlas9/Atlas12} codes \citep{Kurucz93,Kurucz05}. A detailed description is provided by \citet{Aschenbrenneretal23}.} 
We adopted state-of-the-art model atoms; in Table~\ref{table:modelatoms}, for each element, the ions considered are listed along with the number of explicit terms (plus superlevels), radiative bound-bound transitions, and references. All model atoms are completed by the ground term of the next-highest ionisation stage (this is not indicated in the table).

Most of the model atoms have been used in other studies. We note that the model atom for \ion{O}{ii/iii}, which was briefly introduced by \citet{Aschenbrenneretal23}, was slightly extended for the present work, with a 2$s^2$2$p^2$($^1$D)4$f$H\,$^2$[5]\degr~term added to \ion{O}{ii}. In addition, new models for \ion{N}{iii/iv} were assembled. In summary, the level energies for both ions were adopted from \citet{Moore93}. They were combined into 47 LS-coupled (Russell-Saunders coupling) doublet and quartet terms up to the principal quantum number $n$\,=\,7 for \ion{N}{iii}, with the levels for $n$\,=8, 9, and 10 combined into three superlevels. The same strategy was followed for \ion{N}{iv}, which yielded 84 LS-coupled singlet and triplet terms, and three superlevels for $n$\,=8, 9, and 10 in each spin system. The oscillator strengths and photoionisation cross-sections were for the most part adopted from the Opacity Project \citep[OP, e.g.][]{Seatonetal94}, with the data described by \citet{Fernleyetal99} for \ion{N}{iii} and \citet{Tullyetal90} for \ion{N}{iv}. Some improved oscillator strengths were taken from \citet{Wieseetal96} and \citet{FFT04}. Photoionisation data missing in the OP were treated as hydrogenic. Electron impact-excitation data for a large number of transitions were adopted from the ab initio calculations of \citet{Liangetal12} for \ion{N}{iii} and of \citet{Fernandez-Mencheroetal17} for \ion{N}{iv}. Missing data were calculated using Van Regemorter's formula \citep{vanRegemorter62} for radiatively permitted transitions or Allen's formula \citep{Allen73} for forbidden transitions. All collisional ionisation data were calculated via the Seaton formula \citep{Seaton62}, using OP photoionisation threshold cross-sections or hydrogenic values. For the spectrum synthesis of \ion{N}{iii/iv} with {\sc Surface}, we employed oscillator strengths taken from calculations based on the multi-configuration Hartree-Fock method \citep{FFT04} and from Kurucz\footnote{\tt \url{http://kurucz.harvard.edu/atoms.html}}.

\begin{table}
\caption{Model atoms for non-LTE calculations with {\sc Detail}.}  
\label{table:modelatoms}      
\centering                        
{\small
\begin{tabular}{llll}        
\hline\hline
Ion                         & Terms           & Transitions    & Reference \\ \hline
\ion{H}{i}                  & 20              & 190            & {[}1{]}   \\
\ion{He}{i/ii}              & 29+6/20         & 162/190        & {[}2{]}   \\
\ion{C}{ii/iii/iv}          & 68/70/53        & 425/373/319    & {[}3{]}   \\
\ion{N}{i/ii/iii/iv}        & 89/77/47+3/84+6 & 668/462/410/1123 & {[}4{]}   \\
O\,{\sc i/ii/iii}           & 51/177+2/132+2  & 243/2560/1515  & {[}5{]}   \\
Ne\,{\sc i/ii}              & 153/78          & 952/992        & {[}6{]}   \\
Mg\,{\sc ii}                & 37              & 236            & {[}7{]}   \\
Al\,{\sc ii/iii}            & 54+6/46+1       & 378/272        & {[}8{]}   \\
Si\,{\sc ii/iii/iv}         & 52+3/68+4/33+2  & 357/572/242    & {[}9{]}   \\
S\,{\sc ii/iii}             & 78/21           & 302/34         & {[}10{]}  \\
Fe\,{\sc ii/iii/iv}         & 265/60+46/65+70 & 2887/2446/2094 & {[}12{]}\\\hline
\end{tabular}
\tablebib{[1] \cite{PrBu04}; [2] \cite{przybilla05}; [3]~\citet{NiPr06,NiPr08}; [4] \cite{PrBu01}, Przybilla \& Aschenbrenner (in prep.); [5] \cite{Przybillaetal00}, Przybilla \& Butler (in prep.); [6]~\cite{MoBu08}; [7] \cite{Przybillaetal01a}; [8] Przybilla (in prep.); [9] Przybilla \& Butler (in prep.); [10] \citet{Vranckenetal96}, updated; [11] Butler (in prep.); [12] \cite{Becker98}, \cite{Moreletal06}.
}}
\end{table}

{This hybrid non-LTE approach has previously been used to investigate stellar parameters and abundances of a variety of hot stars, with the complexity of the models increasing over the years. Examples encompass slowly rotating single OB main-sequence stars \citep[e.g.][]{NiPr07,NiPr12,Aschenbrenneretal23}, massive B-type stars of non-standard chemical composition \citep[e.g.][]{Irrgangetal10,Irrgangetal22,Przybillaetal21}, and massive B-type supergiants \citep[e.g.][]{Wessmayeretal22,Wessmayeretal24}.} In this work, the method was applied to a multiple massive star system and to faster rotating stars than investigated before.

\subsection{Composite spectra}
{
A key part of our analysis is a normalised model spectrum, $f_{\mathrm{comp}}$,
composed of the individual stellar components, to fit the observed spectrum.
We generalised Eq. (1) of \citet{Irrgangetal14} from binary to multi-star systems:
\begin{equation}
f_{\mathrm{comp}} = \frac{\sum_i w_i f_{\mathrm{cont},i} f_i}{\sum_i w_i f_{\mathrm{cont},i}}\,,
\label{eq:composite_spectrum}
\end{equation}
where $f_i$ is the normalised flux of the $i$-th star, $f_{\mathrm{cont},i}$ the continuum flux of the $i$-th star,
and $w_i$ a weight factor corresponding to the effective surface area. Potential binary interactions may also be considered via modifications of the latter.
}
\begin{figure*}[ht]
    \centering
    \includegraphics[width=0.97\hsize]{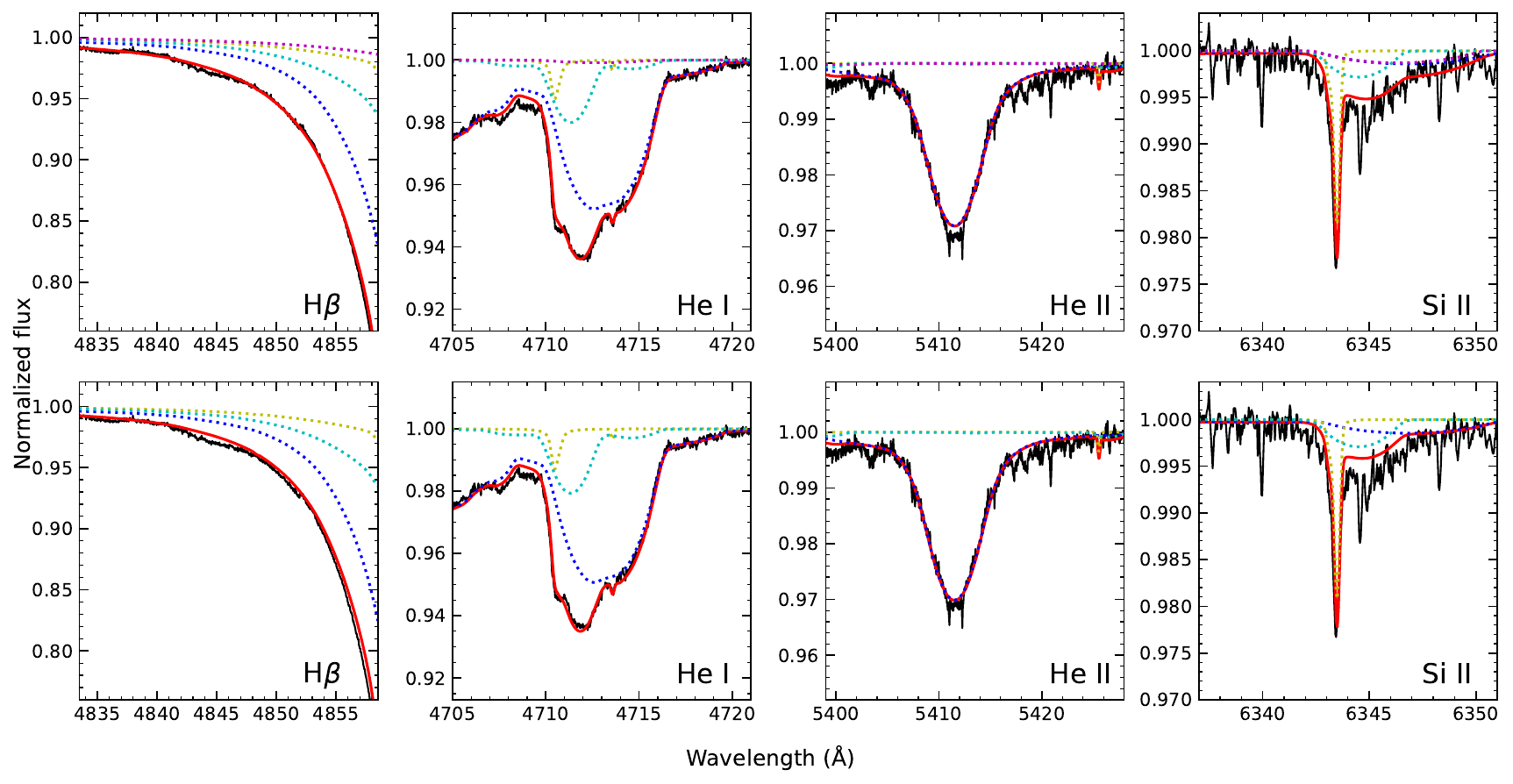}
    \caption{Comparison between the observed spectrum (black) and the synthetic model spectrum (red). The weighted individual contributions of HD~37061\,Aa, Ab, B, and C are shown in dotted blue, yellow, magenta, and cyan, respectively. \textit{Top row:} The model contains the spectral contributions of all four stars. \textit{Bottom row:} The model excluding the spectral contribution of HD~37061\,B.}
    \label{figure:components}
\end{figure*}

\section{Spectral analysis}\label{section:spectral_analysis}
We shifted the observed spectrum into the
rest frame of the primary star (HD~37061\,Aa) by measuring the Doppler shift of the isolated
\ion{He}{ii}, \ion{O}{iii}, and \ion{Si}{iv} lines. In order to fit individual observed spectral lines, we had to include the contribution of all stars simultaneously.
To do so, a grid of model spectra was computed for each star, as was done in our previous work on single stars.
This allowed us to treat each star individually.
We employed self-written Python programs to find the best-fitting grid parameters for all stars by minimising $\chi^2$  using the downhill simplex algorithm \citep{NeMe65}.
At each optimisation step, the free parameters of the model grids were linearly interpolated
to get the individual spectra, $f_i$ (as well as the continuum fluxes, $f_{\mathrm{cont},i}$),
and then the composite spectrum was created according to Eq.~\eqref{eq:composite_spectrum}.
While we could have treated the weights $w_i$ as free parameters, we set them to the square radii of the stars.

In addition, a Doppler shift had to be applied to the spectra of the secondary stars. While the Doppler shift for HD~37061\,Ab and HD~37061\,C can be determined from their absorption lines,
the faintest object in the system, HD~37061\,B, shows no spectral features, due to its low contributed flux and fast rotation.
The object has a separation of $\SI{195 \pm 4}{AU}$ \citep{GravityCol1aboration18} from the primary. Therefore, we assumed a low orbital velocity and shifted the spectrum of HD~37061\,B into the rest frame of the nebula (determined from H$\alpha$ and the forbidden \ion{N}{ii} emission lines in the spectrum).

The contributions of the individual components to the formation of the composite spectrum are visualised for a few examples in Fig.~\ref{figure:components}. Summed up approprietly, these led to a tight fit of the observed spectral features.
The effects from neglecting the flux of HD~37061\,B are also shown in Fig.~\ref{figure:components}. The hydrogen lines of the model become slightly too narrow, and the strengths of helium and most metal lines slightly increase. In the red part of the spectrum, we can observe spectral lines where a decent contribution of the line strength apparently stems from HD~37061\,B (see the \ion{Si}{II} line in Fig.~\ref{figure:components}).

\subsection{Atmospheric parameter and abundance determination}
The atmospheric parameters were determined via an analysis of the spectral lines of multiple ionisation stages from nine different chemical species (C, N, O, Ne, Mg, Al, Si, S, and Fe) and an analysis of neutral and single ionised helium lines and the wings of hydrogen Balmer lines.
These parameters are the effective temperature, $T_{\mathrm{eff}}$, surface gravity, $\log g$, helium number fraction, $y$, microturbulent velocity, $\xi$, projected rotational velocity, $\varv \sin i$, and macroturbulence, $\zeta$, as well as the elemental abundances $\varepsilon\left(X\right)$\,=\,$\log\left(X/\mathrm{H}\right)$\,+\,12.
We derived the parameters on the basis of spectrum synthesis, aiming to simultaneously reproduce detailed line profiles of all stellar components, spanning the entire observed spectral range. 
An iterative approach was used to overcome ambiguities due to correlations between the parameters of individual stars (e.g. $T_{\mathrm{eff}}$ and $\log g$) as well as correlations between the different stars (e.g. element abundances of blended lines), until all parameters were constrained in a single global solution.

\subsubsection{Effective temperature and surface gravity}
We used different chemical elements and ionisation stages to determine the $T_{\mathrm{eff}}$ and $\log g$ of the individual stars. For HD~37061\,Aa, the ionisation equilibria of
\ion{He}{i/ii}, \ion{O}{ii/iii}, and \ion{Si}{iii/iv} were established as temperature indicators, and the surface gravity was adjusted such that the \ion{He}{ii} lines were reproduced. This decision was made due to the gravity-sensitive line strength and the single contribution by HD~37061\,Aa to the \ion{He}{ii} lines (see Fig.~\ref{figure:components}). In the case of HD~37061\,Ab, the ionisation equilibria of \ion{O}{i/ii} and \ion{Si}{ii/iii} were considered as temperature indicators, and the Stark-broadened wings of the hydrogen Balmer lines were used as surface gravity indicators. For HD~37061\,C, we used the ionisation equilibria of \ion{C}{ii/iii} and \ion{O}{i/ii} together with the wings of the hydrogen Balmer lines.

\subsubsection{Projected rotational velocity and macroturbulence}
Both the projected rotational velocity and the macroturbulence were derived simultaneously by fitting the line profiles of isolated metal lines and the wings of blended metal, as well as helium lines. The synthetic models were broadened with rotational and radial-tangential macroturbulence profiles \citep{Gray05} and a Gaussian instrumental profile appropriate for the resolving power of the spectrograph.

\begin{table*}[th]
\caption{Stellar parameters of the stars in the HD\,37061 system.}
\label{tab:stellar_parameters}
\centering   
{\small
\setlength{\tabcolsep}{1.6mm}
\begin{tabular}{lr@{\hspace{0.1mm}}rrcrrrlccrrrrrr}
\hline\hline
Object     &       & $T_{\mathrm{eff}}$ & $\log g$\tablefootmark{a} & $y$ & $\xi$                & $\varv  \sin i$          & $\zeta$              & $R_V$ & $E\left(B-V\right)$ & $B.C.$     & $M_{\mathrm{evol}}$ & $R$         & $\log L/L_\sun$ & $\log \tau_\mathrm{evol}$ & $d_{\mathrm{spec}}$ & $d_{\mathrm{Gaia}}$\tablefootmark{b} \\ \cline{6-8}
&       & kK                 & (cgs)      & {{\tiny by number}}& \multicolumn{3}{c}{$\mathrm{km\,s}^{-1}$} &       & mag                 & mag      & $M_{\odot}$         & $R_{\odot}$ &  & yr & pc                  & pc                  \\ \hline
HD\,37061 &  &  &   &  &  &  &  & 4.75 & 0.51 & &  &  &  &  & 415 & 416 \\
& $\pm$ &  &   &  & &  &  & 0.10 & 0.02 &  &  & & &  & 29 & 10 \\
HD\,37061\,Aa &  & 31.1  & 4.15  & 0.100 & 6 & 190 & 20 & &  & $-$2.94 & 16.4 & 5.7 & 4.43 & 6.5 & & \\
& $\pm$ & 0.5 & 0.05  & 0.006 & 2 & 10 & 3 &  &  & ~~\,0.04 & 0.4 & 0.3 & 0.04 & 0.2 & & \\
HD\,37061\,Ab &  & 15.7  & 4.28  & 0.114 & 1 & 8 & 2 & & & $-$1.32 & 4.4 & 2.5 & 2.54 & $>6.0$ &  &  \\
& $\pm$ & 0.5 & 0.10  & 0.010 & 2 & 3 & 1 & & & ~~\,0.08 & 0.2 & 0.3 & 0.08 &  &  &  \\
HD\,37061\,C &  & 22.3  & 4.25  & 0.101 & 2 & 104 & 20 & & & $-$2.21 & 8.1 & 3.6 & 3.45 & 6.7 &  &  \\
& $\pm$ & 0.5 & 0.10  & 0.011 & 2 & 10 & 5 & & & ~~\,0.06 & 0.4 & 0.4 & 0.08 & 1.0 &  &  \\
\hline
HD\,37061\,B &  & 14.0 & 4.28 & 0.100 & 2 & 180 & 20 & & & $-$1.02 & 3.8 & 2.3 & 2.27 &  &  &  \\
\hline
\end{tabular}
\tablefoot{The first row lists parameters for the whole system. Uncertainties are 1$\sigma$ values.
\tablefoottext{a}{{The values of $\log g$ have not been corrected for centrifugal acceleration.}}
\tablefoottext{b}{\cite{Gaia2016,Gaia2023}.}
}}
\end{table*}

\begin{table*}[th]
\caption{Metal abundances $\varepsilon (X)$\,=\,$\log (X/\mathrm{H})+12$ (by number) and metallicity, $Z$ (by mass) of the individual objects.}
\label{tab:abundances}
\centering   
{\small
\setlength{\tabcolsep}{1.5mm}
\begin{tabular}{ll@{\hspace{0.1mm}}lllllllllc}
\hline\hline
Object & & C & N & O & Ne & Mg  & Al & Si & S & Fe & $Z$  \\ \hline
HD 37061 Aa & & 8.30 (4)  & 7.74 (4) & 8.65 (7) & 8.04 (2) & 7.53 (1) & 6.38 (1) & 7.47 (5) & 7.09 (1) & ... & 0.012\\
& $\pm$ & 0.07 & 0.07 & 0.06 & 0.03 & 0.05  & 0.07 & 0.05 & 0.10 & ... & 0.001\\
HD 37061 Ab & & 8.46 (1)  & ... & 8.71 (3) & 8.11 (2) & 7.50 (2)  & 6.23 (1) & 7.56 (5) & 7.20 (4) & 7.48 (2) & 0.013\\
& $\pm$ & 0.10  & ... & 0.06 & 0.03 & 0.07  & 0.10 & 0.02 & 0.05 & 0.04 & 0.001\\
HD 37061 C & & 8.44 (2) & 7.81 (4) & 8.75 (5) & 8.12 (2) & 7.50 (3)  & 6.24 (1) & 7.54 (2) & ... & 7.53 (1) & 0.014\\
& $\pm$ & 0.04 & 0.08 & 0.03 & 0.03 & 0.05  & 0.10 & 0.02 & ... & 0.10 & 0.001\\[-2mm]
\\ \hline
Messier~43$^a$ & & ... & 7.80 & 8.57 & ... & ... & ... & ... & 6.97 & ... & ...\\
& $\pm$          & ... & 0.04 & 0.05 & ... & ... & ... & ... & 0.03 & ... & ...\\[-2mm]
\\ \hline
CAS~$^{b,c}$       &       & 8.35      & 7.79      & 8.76      & 8.09      & 7.56      & 6.30     & 7.50      & 7.14     & 7.52 & 0.014\\
& $\pm$ & 0.04      & 0.04      & 0.05      & 0.05      & 0.05      & 0.07     & 0.06   & 0.06   & 0.03 & 0.002\\ \hline
\end{tabular}
\tablefoot{Uncertainties are 1$\sigma$ values from the line-to-line scatter. The numbers in parentheses quantify the
analysed lines. Missing values are indicated by `...'.~~~ $^{(a)}$~\citet{Simon-Diazetal11}~~~$^{(b)}$~\citet{NiPr12}~~~$^{(c)}$~\citet{Przybillaetal13}}}
\end{table*}

\subsubsection{Microturbulence and elemental abundances}
When analysing single-star spectra, the microturbulence is usually adjusted such that the individual line abundances become independent of the line strengths (measured by means of the equivalent width). Since almost all the lines in the spectrum of HD~37061 are blends of the individual stellar components, this is more difficult. Therefore, we instead adjusted the microturbulence such that the variance of the individual line abundances was minimised. Once all of the previously mentioned atmospheric parameters were fixed, we derived the abundance values. First we determined individual line-by-line abundances, and then we calculated the final abundance values and uncertainties as the statistical mean and $1\sigma$ standard deviation. In addition to the statistical errors, systematic uncertainties had to be accounted for because of the uncertainties of the atmospheric parameters, the setting of the continuum and the uncertainties of the atomic data used in the non-LTE model atoms. These typically amount to about 0.10 to 0.15\,dex \citep[e.g.][]{Przybillaetal00,Przybillaetal01b,NiPr08} and are likely a bit larger here. This is because we set the microturbulence uncertainty to $\SI{\pm 2}{\kilo\meter\per\second}$ as a conservative estimate based on the resolution of our grids, as the number of available lines for determining the abundances, and thus the microturbulence, is more restricted than usual. Modifying the microturbulence within this boundary can change the mean of the derived abundances by $\sim$\,0.1\,dex while doubling the (logarithmic) standard deviation.

\subsubsection{SED fit}\label{subsection:SED_fit}
The observed SED was constructed from IUE spectrophotometry as well as Johnson $UBV$ \citep{UBVphotometry2002} and 2MASS $JHK$ \citep{Cutrietal03} photometry. For the model, we used the {\sc Atlas9} model fluxes of the individual stars, as they can be compared directly with low-resolution observations without any further processing. The model SED is the sum of the individual fluxes, weighted by the squared radii. To account for interstellar extinction, the reddening law of \citet{Fitzpatrick99} was used, parameterised by the colour excess $E(B-V)$ and the ratio of total-to-selective extinction, $R_V$\,=\,$A_V/E(B-V)$.

\subsection{Fundamental parameters}
The stellar evolutionary masses, $M_\mathrm{evol}$, were found by comparing the position of the stars in the spectroscopic Hertzsprung-Russell diagram \citep[sHRD,][]{LaKu14} relative to the Geneva evolutionary tracks \citep{Ekstroemetal12} for rotating stars ($\varv/\varv_{\text{crit}}=0.4$) at solar metallicity. In the case of the apparently slowly rotating star HD~37061\,Ab, we used the Ekström et al.~tracks for non-rotating stars. The radii were derived from the definition of the surface gravity $g$\,=\,$G M / R^2$, and then the luminosity was derived by applying the Stefan-Boltzmann law $L$\,=\,$4\pi\sigma R^2 T_{\mathrm{eff}}^4$, with $\sigma$ being the Stefan-Boltzmann constant. Evolutionary ages, $\tau_\mathrm{evol}$, were derived by comparing the star positions with isochrones based on the Ekström et al. tracks.   

\subsection{Spectroscopic distances}
The stellar flux of an unresolved multi-star system measured at Earth, $f_{\lambda}$\,=$\sum_i$\,$F_{i,\lambda} \cdot (\pi R_i^2 / d^2),$ depends on the astrophysical flux emitted at the surface of each star, $F_{i,\lambda}$, and the distance, $d$, to the system, as well as the individual radii, $R_i$. The observed magnitude, $m_X$, measured with a normalised filter transmission profile, $T_X(\lambda)$, can then be expressed as
\begin{equation}
\begin{split}
m_X & = -2.5 \log \left( \int f_{\lambda} T_X(\lambda) \mathrm{d}\lambda \right) + zp\\
    & = -2.5 \log \left( \int \left( \sum_i \frac{F_{i,\lambda} R_i^2}{\si{\textit{R}_\odot}^2}\right) T_X(\lambda) \mathrm{d}\lambda \right) + zp \\
    & ~~~~ - 2.5 \log \left( \frac{\pi \si{\textit{R}_\odot}^2}{d^2}\right)\,,
\label{eq:mag_observed}
\end{split}
\end{equation}
with $zp$ being the zero point of the given filter, $X$.
Defining the synthetic magnitude, $\Bar{m}_X$, as
\begin{equation}
\Bar{m}_X = -2.5 \log \left( \int \left( \sum_i \frac{F_{i,\lambda}R_i^2}{\si{\textit{R}_\odot}^2} \right) T_X(\lambda) \mathrm{d}\lambda \right) + zp\,
\label{eq:mag_syn}
\end{equation}
{and inserting {Eq.}~\eqref{eq:mag_syn} into {Eq.}~\eqref{eq:mag_observed}, one can solve for the distance:}
\begin{equation}
d_{\mathrm{spec}} = 4\cdot 10^{0.2(m_X-\Bar{m}_X)-8}\,\mathrm{pc}\,.
\label{eq:dspec}
\end{equation}
{In {Eq.}~\eqref{eq:dspec}, we rewrote the constants in parsec:$\sqrt{\pi}\si{\textit{R}_\odot}=4\cdot 10^{-8}\si{pc}$.
Based on the derived atmospheric parameters, the synthetic magnitude can be calculated from the extinction-corrected (Sect. \ref{subsection:SED_fit}) predicted flux using {Eq.}~\eqref{eq:mag_syn}. Since the calculated distance depends solely on
spectroscopically derived parameters, it is called the spectroscopic distance.}
In this work we used the $d_\mathrm{spec}$ derived from the $V$ magnitude.

\section{Results}\label{section:results}
The stellar parameters derived are listed in Table~\ref{tab:stellar_parameters}: the object's name, effective temperature, surface gravity, surface helium abundance by number, microturbulence velocity, projected rotational velocity, macroturbulence velocity, colour excess, ratio of total-to-selective extinction, bolometric correction, evolutionary mass, radius, luminosity, evolutionary age, spectroscopic distance (calculated using Eq.~\eqref{eq:dspec}), and the \textit{Gaia} DR3 distance. The object HD~37061\,B shows no absorption lines in the spectrum; however, the star has noticeable effects on the continuum. We set $T_{\mathrm{eff}}$ and $\log g$ such that the measured magnitude $K$\,=\,8.7\,mag \citep{GravityCol1aboration18} is reproduced. Although other parameters cannot be determined, we can conclude that HD~37061\,B is a fast-rotating star. Otherwise, absorption lines would be visible (e.g. in the \ion{O}{i} $\lambda\lambda$7772/74/75\,{\AA} triplet). Neglecting the flux contribution of this object would introduce systematic effects to other derived parameters.
An increase of $\sim$\,$\SI{0.05}{dex}$ in $\log g$ would be required for the two other secondary stars to reproduce the wings of the Balmer lines. The analysis of the helium and metal lines would lead to overall slightly lower abundances.

\begin{figure*}[ht]
    \centering
    \includegraphics[width=0.89\hsize]{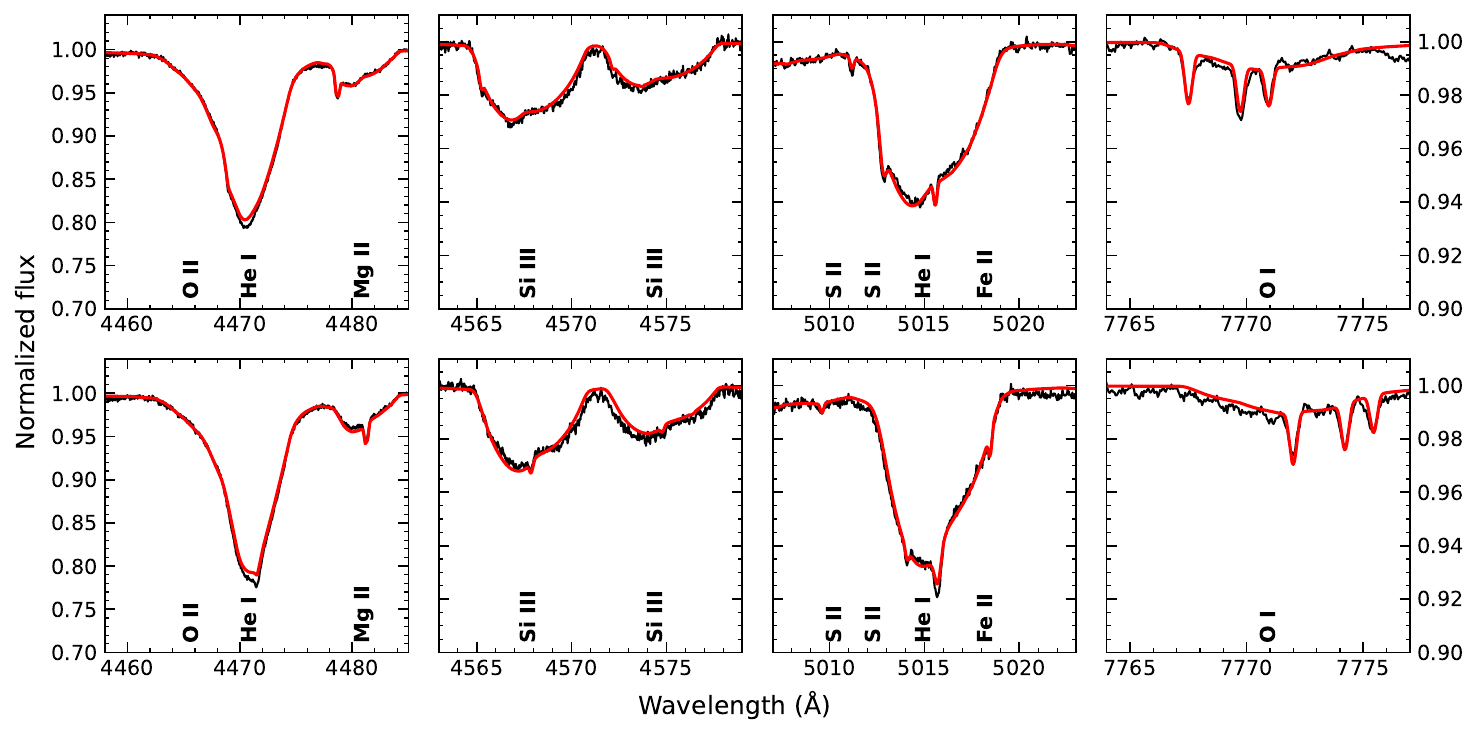}
    \caption{Detailed comparison of the helium and metal lines of the observed spectrum (black) with the global best-fitting solution (red) at different times. The spectra are shifted into the rest frame of the primary star (HD~37061\,Aa). The plots in the first column have a different scaled flux axis. \textit{Top row:} Comparison with the spectrum used for the analysis, observed on March 8, 2007. \textit{Bottom row:} Comparison with a spectrum observed on January 12, 2006. The only difference between the models in the top and bottom row is the radial velocity shift of each stellar component.}
    \label{figure:epochs}
\end{figure*}

The mean abundances of the chemical elements analysed, the uncertainties and number of spectral lines analysed, and the metallicities are listed in Table~\ref{tab:abundances}. While we were able to determine most abundances in the stars, each object is missing one element due to blends or absorption lines that are too weak.
The elemental abundances are compatible with each other. They agree within the mutual uncertainties with the cosmic abundance standard (CAS) derived from massive stars in the solar neighbourhood \citep{NiPr12,Przybillaetal13}, and also with the abundances of the subset of massive stars in Ori OB1 \citep{NiSi11}. In particular, the helium abundance of the magnetic star HD~37061\,C is consistent with standard values despite its position in the $T_\mathrm{eff}$ range,
where magnetic B stars can show He-strong spectral characteristics \citep[e.g.][]{Smith96}. The CNO abundances available for the two massive members of HD~37061 are also consistent with pristine values, locating them at the beginning of the path of mixing signatures in the N/O--N/C diagram \citep{Przybillaetal10,Maederetal14}, which is not unexpected for objects that have evolved only slightly from the zero-age main sequence.

\begin{figure}
    \centering
    \includegraphics[width=0.97\hsize]{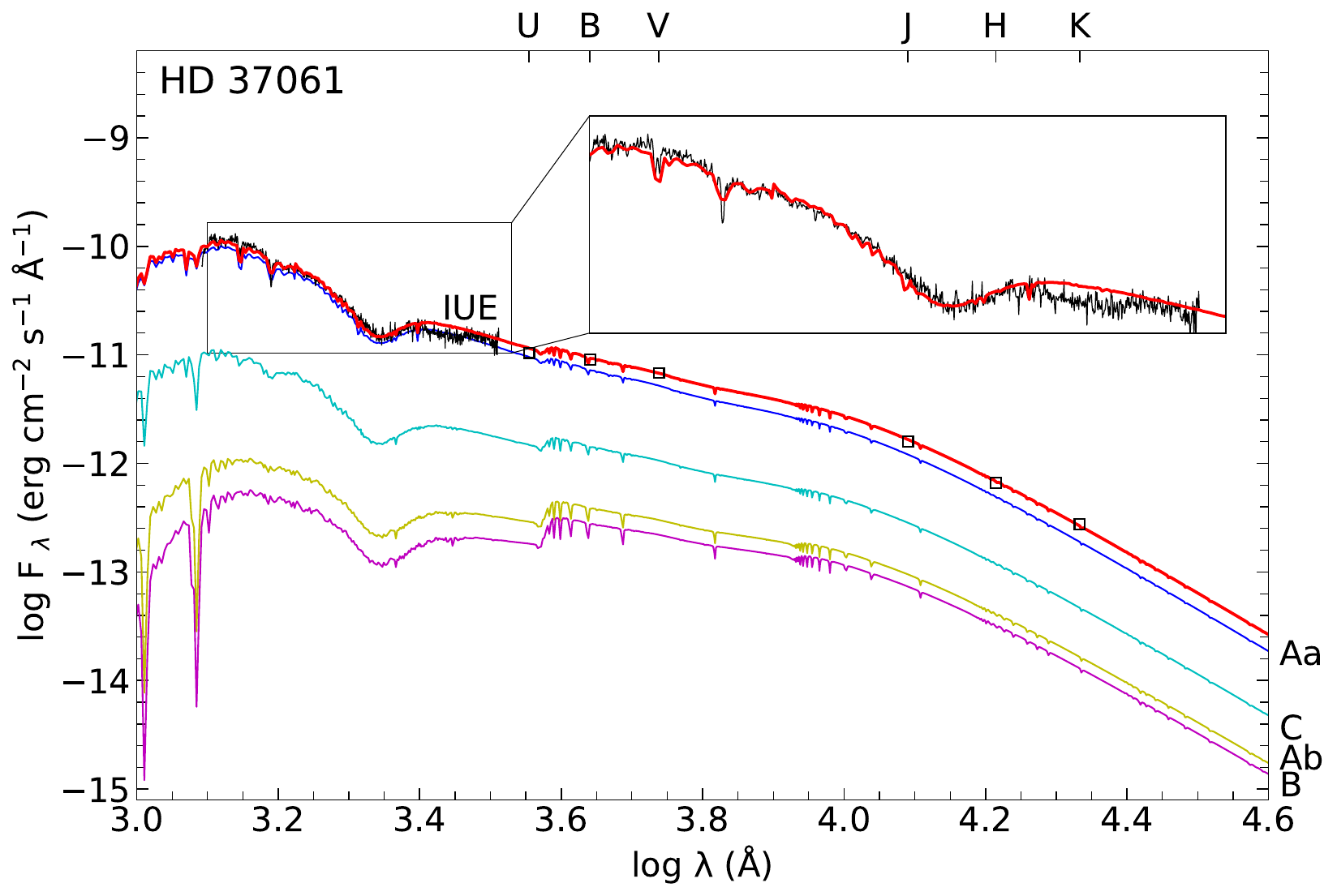}

    \includegraphics[width=0.97\hsize]{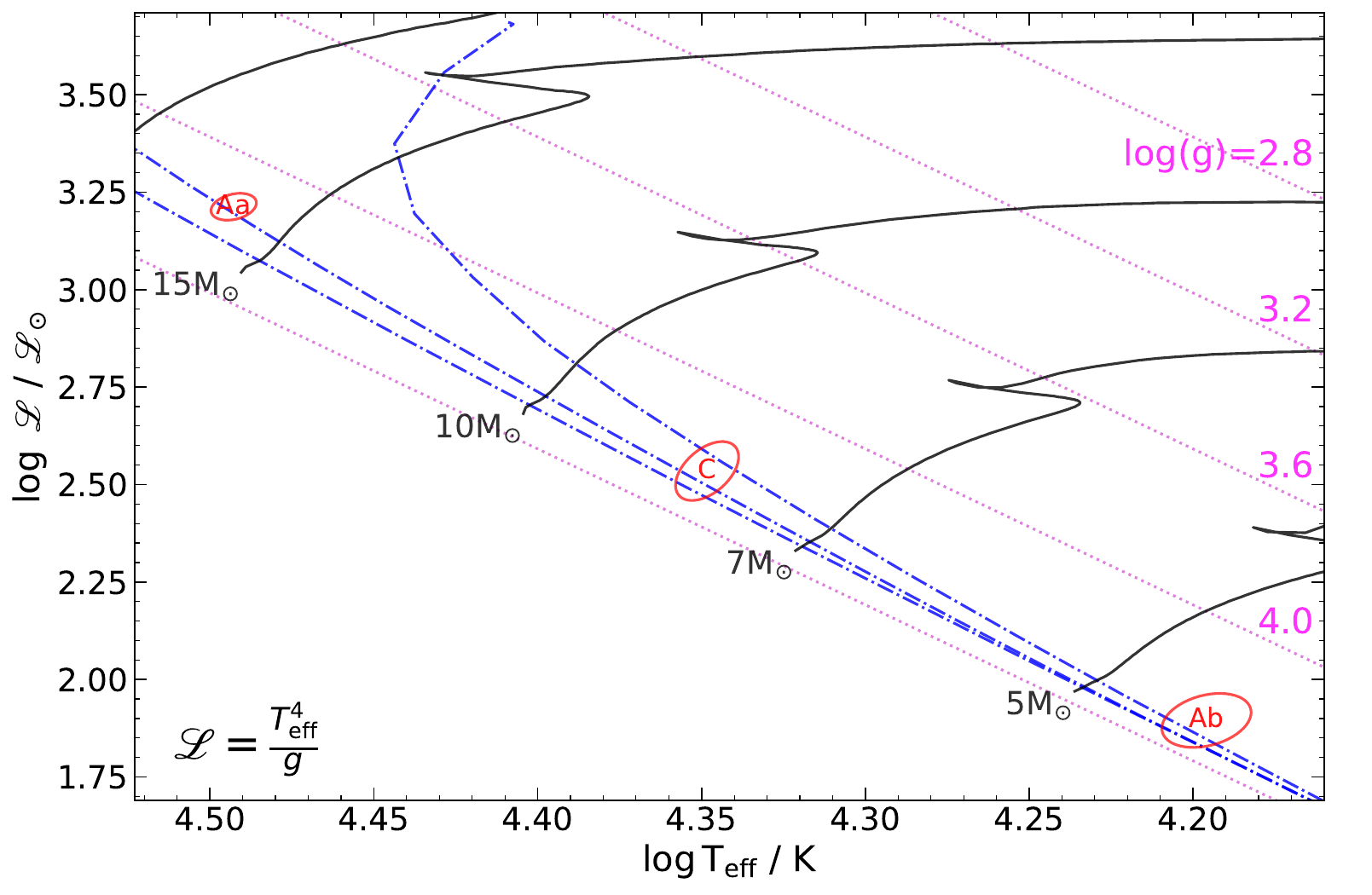}
    \caption{Comparison of the total SED of the HD~37061 system and the location of the individual stars in the sHRD. \textit{Upper panel:} SED fit of the reddened {\sc Atlas9} flux (red) to observed UV spectra (black) and photometric measurements (black squares). The inset zooms in on the UV range. The individual contributions of HD~37061\,Aa, Ab, B, and C are shown in blue, yellow, magenta, and cyan, respectively. \textit{Lower panel:} Position of the stars in the sHRD. The error ellipses outline 1$\sigma$ uncertainties; the stars are labelled. Lines of constant $\log g$ are indicated by dotted magenta lines. The evolutionary tracks of \citet{Ekstroemetal12} for rotating stars  ($v/v_{\text{crit}}=0.4$) with solar metallicity are overlaid in black. The dash-dotted blue lines are isochrones from \citet{Ekstroemetal12} for $\log\tau_{\mathrm{evol}}(\si{yr})=6.0$, $6.5$, and $7.0$.}
    \label{figure:SED_fit}
\end{figure}

We show a comparison between the observed spectrum and our global best-fitting solution in Fig.~\ref{fig:full_fit}. The noticeable differences between the model and observations originate mostly from diffuse interstellar bands (DIBs) and telluric absorption due to O$_2$ and H$_2$O in the Earth's atmosphere. In addition, narrow interstellar absorption lines are visible: the \ion{Ca}{}~H+K lines, the \ion{Na}{}~D lines, and the \ion{He}{i} $\lambda$3889\,{\AA} line originating from the metastable $2s$\,$^3$S level. The spectrum also shows several emission lines from the \ion{H}{ii} region that surrounds HD~37061: H$\alpha$, [\ion{N}{ii}] $\lambda\lambda$6548, 6583\,{\AA,} and [\ion{S}{ii}] $\lambda\lambda$6716, 6731\,{\AA}.
A closer look at selected helium and metal lines is shown in Fig.~\ref{figure:epochs}. As a validation of our solution, we included a comparison with an additional spectrum, observed about a year earlier. The relative movement of the individual components from the orbital motion is clearly visible. By changing only the Doppler shifts, we are able to reproduce this observation as well.

Table \ref{tab:abundances} also includes the nebular abundances of Messier~43 as derived by \citet{Simon-Diazetal11}  from collisionally excited lines.
The nebular nitrogen abundances agree with the stellar ones. However, nebular oxygen and sulphur appear slightly depleted relative to the stellar abundances. This is likely a consequence of the depletion on dust particles, which means that not all the dust has been destroyed yet, and that the metals have not yet been fully incorporated into the gas phase in the \ion{H}{ii} region. Some systematic effects can also be expected to affect the nebular abundances, as reflected by the well-known dichotomy of abundances derived from collisionally excited and recombination lines (see, for example, \citealt{SiSt11} for the case of the Orion nebula, where results from recombination lines required less depletion of metals onto dust).

A fit of the SED, that is, the reproduction of the global energy output of the HD~37061 system, is shown in the upper panel of Fig. \ref{figure:SED_fit}. As with the detailed spectrum, a tight match is found. However, we were unable to achieve a good fit by varying only $R_V$ and $E(B-V)$. We also had to adjust two additional parameters of the \citet{Fitzpatrick99} reddening law: the UV-bump strength was set to $c3$\,=\,2.26 and the far-UV curvature to $c4$\,=\,0.20. These values are lower than the average parameters valid for the Milky Way ($c3$\,=\,3.23, $c4$\,=\,0.41), a situation that can be expected for a sight line that is not dominated by the diffuse ISM but by the material of the remainder of the molecular cloud and the \ion{H}{ii} region. We also note that the reddening parameters for HD~37061 from \citet{Fitzpatrick99} match our values within the mutual uncertainties, despite his treatment of HD~37061 as a single star. The findings of \citet{Valencicetal04} differ more; in particular, their $R_V$\,=\,4.29 is somewhat smaller than our value, and their $c3/R_V$\,=\,0.31 value is significantly smaller. This could be a consequence of them fitting all eight parameters of the Fitzpatrick reddening law and establishing reddening properties by comparing the observed composite SED of HD~37061 with an un-reddened spectral template of a single star that may have differed from the one employed by Fitzpatrick.

In the lower panel of Fig. \ref{figure:SED_fit}, we show the positions of the stars analysed in the sHRD ($\log(\mathscr{L}/\mathscr{L}_{\odot})$ vs $\log T_\mathrm{eff}$, with $\mathscr{L}$\,=\,$T_\mathrm{eff}^4/g$) based solely on the observed atmospheric parameters. The position of the stars relative to the isochrones and evolutionary tracks suggests that HD~37061 is a young system whose individual stars are coeval. The primary component, HD~37061\,Aa, provides the highest sensitivity for the age determination of the system due to its faster evolution.
The typical age of the members of the Orion Nebula cluster is widely agreed to be about 1--2\,Myr \citep[for an overview, see][]{Muenchetal08}. \citet{Simon-Diazetal06} derived an age of 2.5$\pm$0.5\,Myr for the Orion Trapezium stars. Our age determination of 3$^{+2}_{-1}$, Myr is compatible with these values, though somewhat on the high end.

We show a comparison of the stellar parameters of the components of HD~37061 as derived by \citet{Shultzetal19} and from the present work in Table~\ref{table:result_comparison}. There is a remarkable agreement between the values despite the different methods used for their derivation. Shultz et al. used the atmospheric parameters of \citet[][who analysed HD~37061 as a single star]{Simon-Diazetal11} as a starting point and constrained the physical parameters of the system's stars using both orbital models and the same evolutionary tracks that we employed here. Our approach relies on spectroscopic analysis, the use of evolutionary models, and some auxiliary data,
but not on orbital information. Our systematically slightly larger mass, radius, and luminosity values are a consequence of the slightly larger distance and a slightly higher extinction value we employed. This gives us confidence that future applications of our method can be used to characterise binary and multiple stars with a high degree of confidence from a single spectrum. Time-series spectroscopic investigations are still necessary to derive orbital solutions for such systems, but single-epoch spectra from planned future spectroscopic surveys will no longer need to be discarded from the analysis; they can yield much useful data.

\begin{table}
\caption{Comparison of results from \cite{Shultzetal19} and this work.}
\label{table:result_comparison}
\centering  
    \resizebox{\linewidth}{!}{\small\begin{tabular}{lccc} 
\hline\hline                
 Parameter & HD~37061\,Aa & HD~37061\,Ab & HD~37061\,C \\ \hline
 $T_{\mathrm{eff}}$ ($\si{\kilo\kelvin}$) & $\SI{30.5\pm 0.5}{}$ & $\SI{15.2\pm 1.4}{}$ & $\SI{22.2\pm 1.0}{}$ \\
 & $\SI{31.1\pm 0.5}{}$ & $\SI{15.7\pm 0.5}{}$ & $\SI{22.3\pm 0.5}{}$ \\[.7mm]
 $\log g$ & $\SI{4.2\pm 0.1}{}$ & $\SI{4.33\pm 0.01}{}$ & $\SI{4.28\pm 0.02}{}$  \\
 & $\SI{4.15\pm 0.05}{}$ & $\SI{4.28\pm 0.10}{}$ & $\SI{4.25\pm 0.10}{}$  \\[.7mm]
 $M_{\mathrm{evol}}$ ($\si{\Msun}$) & $\SI{14.9\pm 0.5}{}$ & $\SI{3.9\pm 0.7}{}$ & $\SI{7.8\pm 0.7}{}$ \\
 & $\SI{16.4\pm 0.4}{}$ & $\SI{4.4\pm 0.2}{}$ & $\SI{8.1\pm 0.4}{}$ \\[.7mm]
 $R$ ($\si{\textit{R}_\odot}$) & $\SI{5.1\pm 0.3}{}$ & $\SI{2.2\pm 0.2}{}$ & $\SI{3.3\pm 0.2}{}$ \\
 & $\SI{5.7\pm 0.3}{}$ & $\SI{2.5\pm 0.3}{}$ & $\SI{3.6\pm 0.4}{}$ \\[.7mm]
 $\log L$/$\si{\Lsun}$ & $\SI{4.29\pm 0.09}{}$ & $\SI{2.4\pm 0.3}{}$ & $\SI{3.3\pm 0.2}{}$\\
 & $\SI{4.43\pm 0.04}{}$ & $\SI{2.54\pm 0.08}{}$ & $\SI{3.45\pm 0.08}{}$\\[.7mm]
 $\varv \sin i$ ($\si{\kilo\meter\per\second}$) & $\SI{190\pm 10}{}$ & $\SI{10\pm 5}{}$ & $\SI{100\pm 10}{}$ \\
 & $\SI{190\pm 10}{}$ & $\SI{8\pm 3}{}$ & $\SI{104\pm 10}{}$ \\[.7mm]
 $i_{\mathrm{rot}}$ ($^\circ$) & ... & ... & $\SI{38\pm 5}{}$ \\
 & ... & ... & $\SI{39\pm 7}{}$ \\[.7mm]
 $a$ ($\si{AU}$) & ... & ... & $\SI{3.38\pm 0.26}{}$  \\
 & ... & ... & $\SI{3.77\pm 0.12}{}$  \\
 \hline
\end{tabular}}
\tablefoot{In each doublet of values, the upper one is from \citet{Shultzetal19} and the lower is from the present work. We adopted the rotational period and angular separation of HD~37061\,C from \cite{Shultzetal19}. To calculate $a$, we used the \textit{Gaia} DR3 distance.}
\end{table}

Our solution also allows an update of the inclination angle for HD~37061\,C, as the true equatorial rotation velocity can be calculated from the radius and the rotational period, $P$\,=\,1.09478(7)\,d, as determined by Shulz et al. via time-series spectropolarimetry of this magnetic star. It also allows the semi-major axis of the orbit of HD~37061\,C around the HD~37061\,A binary to be reassessed (see also Table~\ref{table:result_comparison}). 

\section{Summary and conclusion}\label{section:summary}
We were able to derive the physical parameters and chemical abundances of three stars in the quadruple system HD~37061. The fourth and faintest star in the system shows no distinct features because of its fast rotation, but contributes noticeable light to the combined continuum of the stars. We modelled the observed composite spectrum by globally fitting the parameters of all stars simultaneously.
For the analysis, we assumed that there were no binary interactions because of the stars' sufficiently large separations, and that each object could therefore be treated as a single star. The combined synthetic spectrum was compared with observations taken about a year apart to verify our approach; we confirm that our approach yielded equally good fits by only varying radial velocities. The derived ages and chemical abundances of the individual stars suggest a common origin and are compatible with typical values for objects in the Orion Nebula region.
A fit to the observed SED yielded a good agreement, and  our calculated spectroscopic distance based on the derived interstellar extinction curve matches the \textit{Gaia} distance.

Our results were derived on the basis of a single high-resolution spectrum and are in remarkable agreement with values derived from a spectroscopic time-series analysis that allowed orbital solutions to be obtained and yielded stellar parameters in an independent way \citep{Shultzetal19}. This provides confidence for future applications of our method when only single observations are available.

\begin{acknowledgements}
The authors are grateful to K. Butler for his advice on atomic data and for many updates and extensions of {\sc Detail} and {\sc Surface}, as well as to A.~Irrgang for additional work on the codes. We further thank A.~Ebenbichler for provision of a DIB linelist.
We also thank our referee, S. Sim\'on-D\'iaz, for a constructive report that helped to improve the paper.
P.A.~acknowledges support of this work by grant of a Ph.D. stipend from the Vice Rectorate for Research of the University of Innsbruck.
Based on observations obtained at the Canada-France-Hawaii Telescope (CFHT) which is operated by the National Research Council (NRC) of Canada, the Institut National des Sciences de l'Univers of the Centre National de la Recherche Scientifique (CNRS) of France, and the University of Hawaii. The observations at the CFHT were performed with care and respect from the summit of Maunakea which is a significant cultural and historic site.
\end{acknowledgements}

%
%

\bibliographystyle{aa}
\bibliography{biblio}

\begin{thebibliography}{86}
\expandafter\ifx\csname natexlab\endcsname\relax\def\natexlab#1{#1}\fi

\bibitem[{{Abt} {et~al.}(1991){Abt}, {Wang}, \& {Cardona}}]{Abtetal91}
{Abt}, H.~A., {Wang}, R., \& {Cardona}, O. 1991, \apj, 367, 155

\bibitem[{{Allen}(1973)}]{Allen73}
{Allen}, C.~W. 1973, {Astrophysical quantities}, 3rd edn. (Athlone Press,
  London)

\bibitem[{{Andersen}(1991)}]{Andersen91}
{Andersen}, J. 1991, \aapr, 3, 91

\bibitem[{{Aschenbrenner} {et~al.}(2023){Aschenbrenner}, {Przybilla}, \&
  {Butler}}]{Aschenbrenneretal23}
{Aschenbrenner}, P., {Przybilla}, N., \& {Butler}, K. 2023, \aap, 671, A36

\bibitem[{{Becker}(1998)}]{Becker98}
{Becker}, S.~R. 1998, ASP Conf.~Ser., 131, 137

\bibitem[{{Blomme} {et~al.}(2022){Blomme}, {Daflon}, {Gebran}, {Herrero},
  {Lobel}, {Mahy}, {Martins}, {Morel}, {Berlanas}, {Blaz{\`e}re}, {Fr{\'e}mat},
  {Gosset}, {Ma{\'\i}z Apell{\'a}niz}, {Santos}, {Semaan},
  {Sim{\'o}n-D{\'\i}az}, {Volpi}, {Holgado}, {Jim{\'e}nez-Esteban}, {Nieva},
  {Przybilla}, {Gilmore}, {Randich}, {Negueruela}, {Prusti}, {Vallenari},
  {Alfaro}, {Bensby}, {Bragaglia}, {Flaccomio}, {Francois}, {Korn},
  {Lanzafame}, {Pancino}, {Smiljanic}, {Bergemann}, {Carraro}, {Franciosini},
  {Gonneau}, {Heiter}, {Hourihane}, {Jofr{\'e}}, {Magrini}, {Morbidelli},
  {Sacco}, {Worley}, \& {Zaggia}}]{Blommeetal22}
{Blomme}, R., {Daflon}, S., {Gebran}, M., {et~al.} 2022, \aap, 661, A120

\bibitem[{{Bragan{\c{c}}a} {et~al.}(2012){Bragan{\c{c}}a}, {Daflon}, {Cunha},
  {Bensby}, {Oey}, \& {Walth}}]{Bragancaetal12}
{Bragan{\c{c}}a}, G.~A., {Daflon}, S., {Cunha}, K., {et~al.} 2012, \aj, 144,
  130

\bibitem[{{Butler} \& {Giddings}(1985)}]{BuGi85}
{Butler}, K. \& {Giddings}, J.~R. 1985, Newsletter of Analysis of Astronomical
  Spectra, 9 (Univ. London)

\bibitem[{{Carneiro} {et~al.}(2019){Carneiro}, {Puls}, {Hoffmann}, {Holgado},
  \& {Sim{\'o}n-D{\'\i}az}}]{Carneiroetal19}
{Carneiro}, L.~P., {Puls}, J., {Hoffmann}, T.~L., {Holgado}, G., \&
  {Sim{\'o}n-D{\'\i}az}, S. 2019, \aap, 623, A3

\bibitem[{{Chini} {et~al.}(2012){Chini}, {Hoffmeister}, {Nasseri}, {Stahl}, \&
  {Zinnecker}}]{Chinietal12}
{Chini}, R., {Hoffmeister}, V.~H., {Nasseri}, A., {Stahl}, O., \& {Zinnecker},
  H. 2012, \mnras, 424, 1925

\bibitem[{{Cutri} {et~al.}(2003){Cutri}, {Skrutskie}, {van Dyk}, {Beichman},
  {Carpenter}, {Chester}, {Cambresy}, {Evans}, {Fowler}, {Gizis}, {Howard},
  {Huchra}, {Jarrett}, {Kopan}, {Kirkpatrick}, {Light}, {Marsh}, {McCallon},
  {Schneider}, {Stiening}, {Sykes}, {Weinberg}, {Wheaton}, {Wheelock}, \&
  {Zacarias}}]{Cutrietal03}
{Cutri}, R.~M., {Skrutskie}, M.~F., {van Dyk}, S., {et~al.} 2003, VizieR Online
  Data Catalog, 2246

\bibitem[{{Ducati}(2002)}]{UBVphotometry2002}
{Ducati}, J.~R. 2002, VizieR Online Data Catalog, 2237

\bibitem[{{Ekstr{\"o}m} {et~al.}(2012){Ekstr{\"o}m}, {Georgy}, {Eggenberger},
  {Meynet}, {Mowlavi}, {Wyttenbach}, {Granada}, {Decressin}, {Hirschi},
  {Frischknecht}, {Charbonnel}, \& {Maeder}}]{Ekstroemetal12}
{Ekstr{\"o}m}, S., {Georgy}, C., {Eggenberger}, P., {et~al.} 2012, \aap, 537,
  A146

\bibitem[{{Fabry} {et~al.}(2021){Fabry}, {Hawcroft}, {Frost}, {Mahy},
  {Marchant}, {Le Bouquin}, \& {Sana}}]{Fabryetal21}
{Fabry}, M., {Hawcroft}, C., {Frost}, A.~J., {et~al.} 2021, \aap, 651, A119

\bibitem[{{Feigelson} {et~al.}(2002){Feigelson}, {Broos}, {Gaffney}, {Garmire},
  {Hillenbrand}, {Pravdo}, {Townsley}, \& {Tsuboi}}]{Feigelsonetal02}
{Feigelson}, E.~D., {Broos}, P., {Gaffney}, James~A., I., {et~al.} 2002, \apj,
  574, 258

\bibitem[{{Fern{\'a}ndez-Menchero} {et~al.}(2017){Fern{\'a}ndez-Menchero},
  {Zatsarinny}, \& {Bartschat}}]{Fernandez-Mencheroetal17}
{Fern{\'a}ndez-Menchero}, L., {Zatsarinny}, O., \& {Bartschat}, K. 2017, J.
  Phys. B, 50, 065203

\bibitem[{{Fernley} {et~al.}(1999){Fernley}, {Hibbert}, {Kingston}, \&
  {Seaton}}]{Fernleyetal99}
{Fernley}, J.~A., {Hibbert}, A., {Kingston}, A.~E., \& {Seaton}, M.~J. 1999, J.
  Phys. B, 32, 5507

\bibitem[{{Fitzpatrick}(1999)}]{Fitzpatrick99}
{Fitzpatrick}, E.~L. 1999, \pasp, 111, 63

\bibitem[{{Froese Fischer} \& {Tachiev}(2004)}]{FFT04}
{Froese Fischer}, C. \& {Tachiev}, G. 2004, At. Data and Nucl. Data Tables, 87,
  1

\bibitem[{{Gaia Collaboration} {et~al.}(2016){Gaia Collaboration}, {Prusti},
  {de Bruijne}, {Brown}, {Vallenari}, {Babusiaux}, {Bailer-Jones}, {Bastian},
  {Biermann}, {Evans}, {Eyer}, {Jansen}, {Jordi}, {Klioner}, {Lammers},
  {Lindegren}, {Luri}, {Mignard}, {Milligan}, {Panem}, {Poinsignon},
  {Pourbaix}, {Randich}, {Sarri}, {Sartoretti}, {Siddiqui}, {Soubiran},
  {Valette}, {van Leeuwen}, {Walton}, {Aerts}, {Arenou}, {Cropper}, {Drimmel},
  {H{\o}g}, {Katz}, {Lattanzi}, {O'Mullane}, {Grebel}, {Holland}, {Huc},
  {Passot}, {Bramante}, {Cacciari}, {Casta{\~n}eda}, {Chaoul}, {Cheek}, {De
  Angeli}, {Fabricius}, {Guerra}, {Hern{\'a}ndez}, {Jean-Antoine-Piccolo},
  {Masana}, {Messineo}, {Mowlavi}, {Nienartowicz}, {Ord{\'o}{\~n}ez-Blanco},
  {Panuzzo}, {Portell}, {Richards}, {Riello}, {Seabroke}, {Tanga},
  {Th{\'e}venin}, {Torra}, {Els}, {Gracia-Abril}, {Comoretto},
  {Garcia-Reinaldos}, {Lock}, {Mercier}, {Altmann}, {Andrae}, {Astraatmadja},
  {Bellas-Velidis}, {Benson}, {Berthier}, {Blomme}, {Busso}, {Carry},
  {Cellino}, {Clementini}, {Cowell}, {Creevey}, {Cuypers}, {Davidson}, {De
  Ridder}, {de Torres}, {Delchambre}, {Dell'Oro}, {Ducourant}, {Fr{\'e}mat},
  {Garc{\'\i}a-Torres}, {Gosset}, {Halbwachs}, {Hambly}, {Harrison}, {Hauser},
  {Hestroffer}, {Hodgkin}, {Huckle}, {Hutton}, {Jasniewicz}, {Jordan},
  {Kontizas}, {Korn}, {Lanzafame}, {Manteiga}, {Moitinho}, {Muinonen},
  {Osinde}, {Pancino}, {Pauwels}, {Petit}, {Recio-Blanco}, {Robin}, {Sarro},
  {Siopis}, {Smith}, {Smith}, {Sozzetti}, {Thuillot}, {van Reeven}, {Viala},
  {Abbas}, {Abreu Aramburu}, {Accart}, {Aguado}, {Allan}, {Allasia},
  {Altavilla}, {{\'A}lvarez}, {Alves}, {Anderson}, {Andrei}, {Anglada Varela},
  {Antiche}, {Antoja}, {Ant{\'o}n}, {Arcay}, {Atzei}, {Ayache}, {Bach},
  {Baker}, {Balaguer-N{\'u}{\~n}ez}, {Barache}, {Barata}, {Barbier}, {Barblan},
  {Baroni}, {Barrado y Navascu{\'e}s}, {Barros}, {Barstow}, {Becciani},
  {Bellazzini}, {Bellei}, {Bello Garc{\'\i}a}, {Belokurov}, {Bendjoya},
  {Berihuete}, {Bianchi}, {Bienaym{\'e}}, {Billebaud}, {Blagorodnova},
  {Blanco-Cuaresma}, {Boch}, {Bombrun}, {Borrachero}, {Bouquillon}, {Bourda},
  {Bouy}, {Bragaglia}, {Breddels}, {Brouillet}, {Br{\"u}semeister},
  {Bucciarelli}, {Budnik}, {Burgess}, {Burgon}, {Burlacu}, {Busonero}, {Buzzi},
  {Caffau}, {Cambras}, {Campbell}, {Cancelliere}, {Cantat-Gaudin}, {Carlucci},
  {Carrasco}, {Castellani}, {Charlot}, {Charnas}, {Charvet}, {Chassat},
  {Chiavassa}, {Clotet}, {Cocozza}, {Collins}, {Collins}, {Costigan}, {Crifo},
  {Cross}, {Crosta}, {Crowley}, {Dafonte}, {Damerdji}, {Dapergolas}, {David},
  {David}, {De Cat}, {de Felice}, {de Laverny}, {De Luise}, {De March}, {de
  Martino}, {de Souza}, {Debosscher}, {del Pozo}, {Delbo}, {Delgado},
  {Delgado}, {di Marco}, {Di Matteo}, {Diakite}, {Distefano}, {Dolding}, {Dos
  Anjos}, {Drazinos}, {Dur{\'a}n}, {Dzigan}, {Ecale}, {Edvardsson}, {Enke},
  {Erdmann}, {Escolar}, {Espina}, {Evans}, {Eynard Bontemps}, {Fabre},
  {Fabrizio}, {Faigler}, {Falc{\~a}o}, {Farr{\`a}s Casas}, {Faye}, {Federici},
  {Fedorets}, {Fern{\'a}ndez-Hern{\'a}ndez}, {Fernique}, {Fienga}, {Figueras},
  {Filippi}, {Findeisen}, {Fonti}, {Fouesneau}, {Fraile}, {Fraser}, {Fuchs},
  {Furnell}, {Gai}, {Galleti}, {Galluccio}, {Garabato}, {Garc{\'\i}a-Sedano},
  {Gar{\'e}}, {Garofalo}, {Garralda}, {Gavras}, {Gerssen}, {Geyer}, {Gilmore},
  {Girona}, {Giuffrida}, {Gomes}, {Gonz{\'a}lez-Marcos},
  {Gonz{\'a}lez-N{\'u}{\~n}ez}, {Gonz{\'a}lez-Vidal}, {Granvik}, {Guerrier},
  {Guillout}, {Guiraud}, {G{\'u}rpide}, {Guti{\'e}rrez-S{\'a}nchez}, {Guy},
  {Haigron}, {Hatzidimitriou}, {Haywood}, {Heiter}, {Helmi}, {Hobbs},
  {Hofmann}, {Holl}, {Holland}, {Hunt}, {Hypki}, {Icardi}, {Irwin}, {Jevardat
  de Fombelle}, {Jofr{\'e}}, {Jonker}, {Jorissen}, {Julbe}, {Karampelas},
  {Kochoska}, {Kohley}, {Kolenberg}, {Kontizas}, {Koposov}, {Kordopatis},
  {Koubsky}, {Kowalczyk}, {Krone-Martins}, {Kudryashova}, {Kull}, {Bachchan},
  {Lacoste-Seris}, {Lanza}, {Lavigne}, {Le Poncin-Lafitte}, {Lebreton},
  {Lebzelter}, {Leccia}, {Leclerc}, {Lecoeur-Taibi}, {Lemaitre}, {Lenhardt},
  {Leroux}, {Liao}, {Licata}, {Lindstr{\o}m}, {Lister}, {Livanou}, {Lobel},
  {L{\"o}ffler}, {L{\'o}pez}, {Lopez-Lozano}, {Lorenz}, {Loureiro},
  {MacDonald}, {Magalh{\~a}es Fernandes}, {Managau}, {Mann}, {Mantelet},
  {Marchal}, {Marchant}, {Marconi}, {Marie}, {Marinoni}, {Marrese},
  {Marschalk{\'o}}, {Marshall}, {Mart{\'\i}n-Fleitas}, {Martino}, {Mary},
  {Matijevi{\v{c}}}, {Mazeh}, {McMillan}, {Messina}, {Mestre}, {Michalik},
  {Millar}, {Miranda}, {Molina}, {Molinaro}, {Molinaro}, {Moln{\'a}r},
  {Moniez}, {Montegriffo}, {Monteiro}, {Mor}, {Mora}, {Morbidelli}, {Morel},
  {Morgenthaler}, {Morley}, {Morris}, {Mulone}, {Muraveva}, {Musella},
  {Narbonne}, {Nelemans}, {Nicastro}, {Noval}, {Ord{\'e}novic},
  {Ordieres-Mer{\'e}}, {Osborne}, {Pagani}, {Pagano}, {Pailler}, {Palacin},
  {Palaversa}, {Parsons}, {Paulsen}, {Pecoraro}, {Pedrosa}, {Pentik{\"a}inen},
  {Pereira}, {Pichon}, {Piersimoni}, {Pineau}, {Plachy}, {Plum}, {Poujoulet},
  {Pr{\v{s}}a}, {Pulone}, {Ragaini}, {Rago}, {Rambaux}, {Ramos-Lerate},
  {Ranalli}, {Rauw}, {Read}, {Regibo}, {Renk}, {Reyl{\'e}}, {Ribeiro},
  {Rimoldini}, {Ripepi}, {Riva}, {Rixon}, {Roelens}, {Romero-G{\'o}mez},
  {Rowell}, {Royer}, {Rudolph}, {Ruiz-Dern}, {Sadowski}, {Sagrist{\`a}
  Sell{\'e}s}, {Sahlmann}, {Salgado}, {Salguero}, {Sarasso}, {Savietto},
  {Schnorhk}, {Schultheis}, {Sciacca}, {Segol}, {Segovia}, {Segransan},
  {Serpell}, {Shih}, {Smareglia}, {Smart}, {Smith}, {Solano}, {Solitro},
  {Sordo}, {Soria Nieto}, {Souchay}, {Spagna}, {Spoto}, {Stampa}, {Steele},
  {Steidelm{\"u}ller}, {Stephenson}, {Stoev}, {Suess}, {S{\"u}veges}, {Surdej},
  {Szabados}, {Szegedi-Elek}, {Tapiador}, {Taris}, {Tauran}, {Taylor},
  {Teixeira}, {Terrett}, {Tingley}, {Trager}, {Turon}, {Ulla}, {Utrilla},
  {Valentini}, {van Elteren}, {Van Hemelryck}, {van Leeuwen}, {Varadi},
  {Vecchiato}, {Veljanoski}, {Via}, {Vicente}, {Vogt}, {Voss}, {Votruba},
  {Voutsinas}, {Walmsley}, {Weiler}, {Weingrill}, {Werner}, {Wevers},
  {Whitehead}, {Wyrzykowski}, {Yoldas}, {{\v{Z}}erjal}, {Zucker}, {Zurbach},
  {Zwitter}, {Alecu}, {Allen}, {Allende Prieto}, {Amorim},
  {Anglada-Escud{\'e}}, {Arsenijevic}, {Azaz}, {Balm}, {Beck}, {Bernstein},
  {Bigot}, {Bijaoui}, {Blasco}, {Bonfigli}, {Bono}, {Boudreault}, {Bressan},
  {Brown}, {Brunet}, {Bunclark}, {Buonanno}, {Butkevich}, {Carret}, {Carrion},
  {Chemin}, {Ch{\'e}reau}, {Corcione}, {Darmigny}, {de Boer}, {de Teodoro}, {de
  Zeeuw}, {Delle Luche}, {Domingues}, {Dubath}, {Fodor}, {Fr{\'e}zouls},
  {Fries}, {Fustes}, {Fyfe}, {Gallardo}, {Gallegos}, {Gardiol}, {Gebran},
  {Gomboc}, {G{\'o}mez}, {Grux}, {Gueguen}, {Heyrovsky}, {Hoar}, {Iannicola},
  {Isasi Parache}, {Janotto}, {Joliet}, {Jonckheere}, {Keil}, {Kim},
  {Klagyivik}, {Klar}, {Knude}, {Kochukhov}, {Kolka}, {Kos}, {Kutka}, {Lainey},
  {LeBouquin}, {Liu}, {Loreggia}, {Makarov}, {Marseille}, {Martayan},
  {Martinez-Rubi}, {Massart}, {Meynadier}, {Mignot}, {Munari}, {Nguyen},
  {Nordlander}, {Ocvirk}, {O'Flaherty}, {Olias Sanz}, {Ortiz}, {Osorio},
  {Oszkiewicz}, {Ouzounis}, {Palmer}, {Park}, {Pasquato}, {Peltzer}, {Peralta},
  {P{\'e}turaud}, {Pieniluoma}, {Pigozzi}, {Poels}, {Prat}, {Prod'homme},
  {Raison}, {Rebordao}, {Risquez}, {Rocca-Volmerange}, {Rosen}, {Ruiz-Fuertes},
  {Russo}, {Sembay}, {Serraller Vizcaino}, {Short}, {Siebert}, {Silva},
  {Sinachopoulos}, {Slezak}, {Soffel}, {Sosnowska}, {Strai{\v{z}}ys}, {ter
  Linden}, {Terrell}, {Theil}, {Tiede}, {Troisi}, {Tsalmantza}, {Tur},
  {Vaccari}, {Vachier}, {Valles}, {Van Hamme}, {Veltz}, {Virtanen}, {Wallut},
  {Wichmann}, {Wilkinson}, {Ziaeepour}, \& {Zschocke}}]{Gaia2016}
{Gaia Collaboration}, {Prusti}, T., {de Bruijne}, J.~H.~J., {et~al.} 2016,
  \aap, 595, A1

\bibitem[{{Gaia Collaboration} {et~al.}(2023){Gaia Collaboration}, {Vallenari},
  {Brown}, {Prusti}, {de Bruijne}, {Arenou}, {Babusiaux}, {Biermann},
  {Creevey}, {Ducourant}, {Evans}, {Eyer}, {Guerra}, {Hutton}, {Jordi},
  {Klioner}, {Lammers}, {Lindegren}, {Luri}, {Mignard}, {Panem}, {Pourbaix},
  {Randich}, {Sartoretti}, {Soubiran}, {Tanga}, {Walton}, {Bailer-Jones},
  {Bastian}, {Drimmel}, {Jansen}, {Katz}, {Lattanzi}, {van Leeuwen}, {Bakker},
  {Cacciari}, {Casta{\~n}eda}, {De Angeli}, {Fabricius}, {Fouesneau},
  {Fr{\'e}mat}, {Galluccio}, {Guerrier}, {Heiter}, {Masana}, {Messineo},
  {Mowlavi}, {Nicolas}, {Nienartowicz}, {Pailler}, {Panuzzo}, {Riclet}, {Roux},
  {Seabroke}, {Sordo}, {Th{\'e}venin}, {Gracia-Abril}, {Portell}, {Teyssier},
  {Altmann}, {Andrae}, {Audard}, {Bellas-Velidis}, {Benson}, {Berthier},
  {Blomme}, {Burgess}, {Busonero}, {Busso}, {C{\'a}novas}, {Carry}, {Cellino},
  {Cheek}, {Clementini}, {Damerdji}, {Davidson}, {de Teodoro}, {Nu{\~n}ez
  Campos}, {Delchambre}, {Dell'Oro}, {Esquej}, {Fern{\'a}ndez-Hern{\'a}ndez},
  {Fraile}, {Garabato}, {Garc{\'\i}a-Lario}, {Gosset}, {Haigron}, {Halbwachs},
  {Hambly}, {Harrison}, {Hern{\'a}ndez}, {Hestroffer}, {Hodgkin}, {Holl},
  {Jan{\ss}en}, {Jevardat de Fombelle}, {Jordan}, {Krone-Martins}, {Lanzafame},
  {L{\"o}ffler}, {Marchal}, {Marrese}, {Moitinho}, {Muinonen}, {Osborne},
  {Pancino}, {Pauwels}, {Recio-Blanco}, {Reyl{\'e}}, {Riello}, {Rimoldini},
  {Roegiers}, {Rybizki}, {Sarro}, {Siopis}, {Smith}, {Sozzetti}, {Utrilla},
  {van Leeuwen}, {Abbas}, {{\'A}brah{\'a}m}, {Abreu Aramburu}, {Aerts},
  {Aguado}, {Ajaj}, {Aldea-Montero}, {Altavilla}, {{\'A}lvarez}, {Alves},
  {Anders}, {Anderson}, {Anglada Varela}, {Antoja}, {Baines}, {Baker},
  {Balaguer-N{\'u}{\~n}ez}, {Balbinot}, {Balog}, {Barache}, {Barbato},
  {Barros}, {Barstow}, {Bartolom{\'e}}, {Bassilana}, {Bauchet}, {Becciani},
  {Bellazzini}, {Berihuete}, {Bernet}, {Bertone}, {Bianchi}, {Binnenfeld},
  {Blanco-Cuaresma}, {Blazere}, {Boch}, {Bombrun}, {Bossini}, {Bouquillon},
  {Bragaglia}, {Bramante}, {Breedt}, {Bressan}, {Brouillet}, {Brugaletta},
  {Bucciarelli}, {Burlacu}, {Butkevich}, {Buzzi}, {Caffau}, {Cancelliere},
  {Cantat-Gaudin}, {Carballo}, {Carlucci}, {Carnerero}, {Carrasco},
  {Casamiquela}, {Castellani}, {Castro-Ginard}, {Chaoul}, {Charlot}, {Chemin},
  {Chiaramida}, {Chiavassa}, {Chornay}, {Comoretto}, {Contursi}, {Cooper},
  {Cornez}, {Cowell}, {Crifo}, {Cropper}, {Crosta}, {Crowley}, {Dafonte},
  {Dapergolas}, {David}, {David}, {de Laverny}, {De Luise}, {De March}, {De
  Ridder}, {de Souza}, {de Torres}, {del Peloso}, {del Pozo}, {Delbo},
  {Delgado}, {Delisle}, {Demouchy}, {Dharmawardena}, {Di Matteo}, {Diakite},
  {Diener}, {Distefano}, {Dolding}, {Edvardsson}, {Enke}, {Fabre}, {Fabrizio},
  {Faigler}, {Fedorets}, {Fernique}, {Fienga}, {Figueras}, {Fournier},
  {Fouron}, {Fragkoudi}, {Gai}, {Garcia-Gutierrez}, {Garcia-Reinaldos},
  {Garc{\'\i}a-Torres}, {Garofalo}, {Gavel}, {Gavras}, {Gerlach}, {Geyer},
  {Giacobbe}, {Gilmore}, {Girona}, {Giuffrida}, {Gomel}, {Gomez},
  {Gonz{\'a}lez-N{\'u}{\~n}ez}, {Gonz{\'a}lez-Santamar{\'\i}a},
  {Gonz{\'a}lez-Vidal}, {Granvik}, {Guillout}, {Guiraud},
  {Guti{\'e}rrez-S{\'a}nchez}, {Guy}, {Hatzidimitriou}, {Hauser}, {Haywood},
  {Helmer}, {Helmi}, {Sarmiento}, {Hidalgo}, {Hilger}, {H{\l}adczuk}, {Hobbs},
  {Holland}, {Huckle}, {Jardine}, {Jasniewicz}, {Jean-Antoine Piccolo},
  {Jim{\'e}nez-Arranz}, {Jorissen}, {Juaristi Campillo}, {Julbe}, {Karbevska},
  {Kervella}, {Khanna}, {Kontizas}, {Kordopatis}, {Korn}, {K{\'o}sp{\'a}l},
  {Kostrzewa-Rutkowska}, {Kruszy{\'n}ska}, {Kun}, {Laizeau}, {Lambert},
  {Lanza}, {Lasne}, {Le Campion}, {Lebreton}, {Lebzelter}, {Leccia}, {Leclerc},
  {Lecoeur-Taibi}, {Liao}, {Licata}, {Lindstr{\o}m}, {Lister}, {Livanou},
  {Lobel}, {Lorca}, {Loup}, {Madrero Pardo}, {Magdaleno Romeo}, {Managau},
  {Mann}, {Manteiga}, {Marchant}, {Marconi}, {Marcos}, {Marcos Santos},
  {Mar{\'\i}n Pina}, {Marinoni}, {Marocco}, {Marshall}, {Martin Polo},
  {Mart{\'\i}n-Fleitas}, {Marton}, {Mary}, {Masip}, {Massari},
  {Mastrobuono-Battisti}, {Mazeh}, {McMillan}, {Messina}, {Michalik}, {Millar},
  {Mints}, {Molina}, {Molinaro}, {Moln{\'a}r}, {Monari}, {Mongui{\'o}},
  {Montegriffo}, {Montero}, {Mor}, {Mora}, {Morbidelli}, {Morel}, {Morris},
  {Muraveva}, {Murphy}, {Musella}, {Nagy}, {Noval}, {Oca{\~n}a}, {Ogden},
  {Ordenovic}, {Osinde}, {Pagani}, {Pagano}, {Palaversa}, {Palicio},
  {Pallas-Quintela}, {Panahi}, {Payne-Wardenaar}, {Pe{\~n}alosa Esteller},
  {Penttil{\"a}}, {Pichon}, {Piersimoni}, {Pineau}, {Plachy}, {Plum}, {Poggio},
  {Pr{\v{s}}a}, {Pulone}, {Racero}, {Ragaini}, {Rainer}, {Raiteri}, {Rambaux},
  {Ramos}, {Ramos-Lerate}, {Re Fiorentin}, {Regibo}, {Richards}, {Rios Diaz},
  {Ripepi}, {Riva}, {Rix}, {Rixon}, {Robichon}, {Robin}, {Robin}, {Roelens},
  {Rogues}, {Rohrbasser}, {Romero-G{\'o}mez}, {Rowell}, {Royer}, {Ruz Mieres},
  {Rybicki}, {Sadowski}, {S{\'a}ez N{\'u}{\~n}ez}, {Sagrist{\`a} Sell{\'e}s},
  {Sahlmann}, {Salguero}, {Samaras}, {Sanchez Gimenez}, {Sanna},
  {Santove{\~n}a}, {Sarasso}, {Schultheis}, {Sciacca}, {Segol}, {Segovia},
  {S{\'e}gransan}, {Semeux}, {Shahaf}, {Siddiqui}, {Siebert}, {Siltala},
  {Silvelo}, {Slezak}, {Slezak}, {Smart}, {Snaith}, {Solano}, {Solitro},
  {Souami}, {Souchay}, {Spagna}, {Spina}, {Spoto}, {Steele},
  {Steidelm{\"u}ller}, {Stephenson}, {S{\"u}veges}, {Surdej}, {Szabados},
  {Szegedi-Elek}, {Taris}, {Taylor}, {Teixeira}, {Tolomei}, {Tonello}, {Torra},
  {Torra}, {Torralba Elipe}, {Trabucchi}, {Tsounis}, {Turon}, {Ulla}, {Unger},
  {Vaillant}, {van Dillen}, {van Reeven}, {Vanel}, {Vecchiato}, {Viala},
  {Vicente}, {Voutsinas}, {Weiler}, {Wevers}, {Wyrzykowski}, {Yoldas}, {Yvard},
  {Zhao}, {Zorec}, {Zucker}, \& {Zwitter}}]{Gaia2023}
{Gaia Collaboration}, {Vallenari}, A., {Brown}, A.~G.~A., {et~al.} 2023, \aap,
  674, A1

\bibitem[{{Giddings}(1981)}]{giddings81}
{Giddings}, J.~R. 1981, PhD thesis, (Univ. London)

\bibitem[{{Gonz{\'a}lez} {et~al.}(2019){Gonz{\'a}lez}, {Briquet}, {Przybilla},
  {Nieva}, {De Cat}, {Saesen}, {Hubrig}, {Thoul}, {P{\'a}pics}, {Palaversa},
  {Naef}, {Neveu-Van Malle}, {J{\"a}rvinen}, {Pollard}, {Kilmartin}, {Mowlavi},
  \& {Butler}}]{Gonzalezetal19}
{Gonz{\'a}lez}, J.~F., {Briquet}, M., {Przybilla}, N., {et~al.} 2019, \aap,
  626, A94

\bibitem[{{Gonz{\'a}lez} {et~al.}(2017){Gonz{\'a}lez}, {Hubrig}, {Przybilla},
  {Carroll}, {Nieva}, {Ilyin}, {J{\"a}rvinen}, {Morel}, {Sch{\"o}ller},
  {Castro}, {Barb{\'a}}, {de Koter}, {Schneider}, {Kholtygin}, {Butler},
  {Veramendi}, {Langer}, \& {BOB Collaboration}}]{Gonzalezetal17}
{Gonz{\'a}lez}, J.~F., {Hubrig}, S., {Przybilla}, N., {et~al.} 2017, \mnras,
  467, 437

\bibitem[{{GRAVITY Collaboration} {et~al.}(2018){GRAVITY Collaboration},
  {Karl}, {Pfuhl}, {Eisenhauer}, {Genzel}, {Grellmann}, {Habibi}, {Abuter},
  {Accardo}, {Amorim}, {Anugu}, {{\'A}vila}, {Benisty}, {Berger}, {Blind},
  {Bonnet}, {Bourget}, {Brandner}, {Brast}, {Buron}, {Caratti O Garatti},
  {Chapron}, {Cl{\'e}net}, {Collin}, {Coud{\'e} Du Foresto}, {de Wit}, {de
  Zeeuw}, {Deen}, {Delplancke-Str{\"o}bele}, {Dembet}, {Derie}, {Dexter},
  {Duvert}, {Ebert}, {Eckart}, {Esselborn}, {F{\'e}dou}, {Finger}, {Garcia},
  {Garcia Dabo}, {Garcia Lopez}, {Gao}, {Gendron}, {Gillessen}, {Gont{\'e}},
  {Gordo}, {Gr{\"o}zinger}, {Guajardo}, {Guieu}, {Haguenauer}, {Hans},
  {Haubois}, {Haug}, {Hau{\ss}mann}, {Henning}, {Hippler}, {Horrobin}, {Huber},
  {Hubert}, {Hubin}, {Jakob}, {Jochum}, {Jocou}, {Kaufer}, {Kellner},
  {Kendrew}, {Kern}, {Kervella}, {Kiekebusch}, {Klein}, {K{\"o}hler}, {Kolb},
  {Kulas}, {Lacour}, {Lapeyr{\`e}re}, {Lazareff}, {Le Bouquin}, {L{\'e}na},
  {Lenzen}, {L{\'e}v{\^e}que}, {Lin}, {Lippa}, {Magnard}, {Mehrgan},
  {M{\'e}rand}, {Moulin}, {M{\"u}ller}, {M{\"u}ller}, {Neumann}, {Oberti},
  {Ott}, {Pallanca}, {Panduro}, {Pasquini}, {Paumard}, {Percheron}, {Perraut},
  {Perrin}, {Pfl{\"u}ger}, {Duc}, {Plewa}, {Popovic}, {Rabien}, {Ram{\'\i}rez},
  {Ramos}, {Rau}, {Riquelme}, {Rodr{\'\i}guez-Coira}, {Rohloff}, {Rosales},
  {Rousset}, {Sanchez-Bermudez}, {Scheithauer}, {Sch{\"o}ller}, {Schuhler},
  {Spyromilio}, {Straub}, {Straubmeier}, {Sturm}, {Suarez}, {Tristram},
  {Ventura}, {Vincent}, {Waisberg}, {Wank}, {Widmann}, {Wieprecht}, {Wiest},
  {Wiezorrek}, {Wittkowski}, {Woillez}, {Wolff}, {Yazici}, {Ziegler}, \&
  {Zins}}]{GravityCol1aboration18}
{GRAVITY Collaboration}, {Karl}, M., {Pfuhl}, O., {et~al.} 2018, \aap, 620,
  A116

\bibitem[{{Gray}(2005)}]{Gray05}
{Gray}, D.~F. 2005, {The Observation and Analysis of Stellar Photospheres}, 3rd
  edn. (Cambridge: Cambridge University Press)

\bibitem[{{Grellmann} {et~al.}(2013){Grellmann}, {Preibisch}, {Ratzka},
  {Kraus}, {Helminiak}, \& {Zinnecker}}]{Grellmannetal13}
{Grellmann}, R., {Preibisch}, T., {Ratzka}, T., {et~al.} 2013, \aap, 550, A82

\bibitem[{{Hadrava}(1995)}]{Hadrava95}
{Hadrava}, P. 1995, \aaps, 114, 393

\bibitem[{{Hunter} {et~al.}(2009){Hunter}, {Brott}, {Langer}, {Lennon},
  {Dufton}, {Howarth}, {Ryans}, {Trundle}, {Evans}, {de Koter}, \&
  {Smartt}}]{Hunteretal09}
{Hunter}, I., {Brott}, I., {Langer}, N., {et~al.} 2009, \aap, 496, 841

\bibitem[{{Irrgang} {et~al.}(2014){Irrgang}, {Przybilla}, {Heber}, {B{\"o}ck},
  {Hanke}, {Nieva}, \& {Butler}}]{Irrgangetal14}
{Irrgang}, A., {Przybilla}, N., {Heber}, U., {et~al.} 2014, \aap, 565, A63

\bibitem[{{Irrgang} {et~al.}(2010){Irrgang}, {Przybilla}, {Heber}, {Nieva}, \&
  {Schuh}}]{Irrgangetal10}
{Irrgang}, A., {Przybilla}, N., {Heber}, U., {Nieva}, M.~F., \& {Schuh}, S.
  2010, \apj, 711, 138

\bibitem[{{Irrgang} {et~al.}(2022){Irrgang}, {Przybilla}, \&
  {Meynet}}]{Irrgangetal22}
{Irrgang}, A., {Przybilla}, N., \& {Meynet}, G. 2022, Nature Astronomy, 6, 1414

\bibitem[{{Johnston} {et~al.}(2019){Johnston}, {Pavlovski}, \&
  {Tkachenko}}]{Johnstonetal19}
{Johnston}, C., {Pavlovski}, K., \& {Tkachenko}, A. 2019, \aap, 628, A25

\bibitem[{{Kurucz}(1993)}]{Kurucz93}
{Kurucz}, R. 1993, CD-ROM No.~13~(Cambridge, Mass.: SAO)

\bibitem[{{Kurucz}(2005)}]{Kurucz05}
{Kurucz}, R.~L. 2005, Mem. Societa Astronomica Italiana Suppl., 8, 14

\bibitem[{{Langer} \& {Kudritzki}(2014)}]{LaKu14}
{Langer}, N. \& {Kudritzki}, R.~P. 2014, \aap, 564, A52

\bibitem[{{Liang} {et~al.}(2012){Liang}, {Badnell}, \& {Zhao}}]{Liangetal12}
{Liang}, G.~Y., {Badnell}, N.~R., \& {Zhao}, G. 2012, \aap, 547, A87

\bibitem[{{Maeder} {et~al.}(2014){Maeder}, {Przybilla}, {Nieva}, {Georgy},
  {Meynet}, {Ekstr{\"o}m}, \& {Eggenberger}}]{Maederetal14}
{Maeder}, A., {Przybilla}, N., {Nieva}, M.~F., {et~al.} 2014, \aap, 565, A39

\bibitem[{{Mahy} {et~al.}(2010){Mahy}, {Rauw}, {Martins}, {Naz{\'e}}, {Gosset},
  {De Becker}, {Sana}, \& {Eenens}}]{Mahyetal10}
{Mahy}, L., {Rauw}, G., {Martins}, F., {et~al.} 2010, \apj, 708, 1537

\bibitem[{{Manset} \& {Donati}(2003)}]{ManDon03}
{Manset}, N. \& {Donati}, J.-F. 2003, Proc. SPIE, 4843, 425

\bibitem[{{Martins} {et~al.}(2017){Martins}, {Sim{\'o}n-D{\'\i}az},
  {Barb{\'a}}, {Gamen}, \& {Ekstr{\"o}m}}]{Martinsetal17}
{Martins}, F., {Sim{\'o}n-D{\'\i}az}, S., {Barb{\'a}}, R.~H., {Gamen}, R.~C.,
  \& {Ekstr{\"o}m}, S. 2017, \aap, 599, A30

\bibitem[{{Moore}(1993)}]{Moore93}
{Moore}, C.~E. 1993, {Tables of Spectra of Hydrogen, Carbon, Nitrogen, and
  Oxygen Atoms and Ions} (Boca Raton, FL: CRC Press)

\bibitem[{{Morel} {et~al.}(2022){Morel}, {Blaz{\`e}re}, {Semaan}, {Gosset},
  {Zorec}, {Fr{\'e}mat}, {Blomme}, {Daflon}, {Lobel}, {Nieva}, {Przybilla},
  {Gebran}, {Herrero}, {Mahy}, {Santos}, {Tautvai{\v{s}}ien{\.{e}}}, {Gilmore},
  {Randich}, {Alfaro}, {Bergemann}, {Carraro}, {Damiani}, {Franciosini},
  {Morbidelli}, {Pancino}, {Worley}, \& {Zaggia}}]{Moreletal22}
{Morel}, T., {Blaz{\`e}re}, A., {Semaan}, T., {et~al.} 2022, \aap, 665, A108

\bibitem[{{Morel} \& {Butler}(2008)}]{MoBu08}
{Morel}, T. \& {Butler}, K. 2008, \aap, 487, 307

\bibitem[{{Morel} {et~al.}(2006){Morel}, {Butler}, {Aerts}, {Neiner}, \&
  {Briquet}}]{Moreletal06}
{Morel}, T., {Butler}, K., {Aerts}, C., {Neiner}, C., \& {Briquet}, M. 2006,
  \aap, 457, 651

\bibitem[{{Morrell} \& {Levato}(1991)}]{MorrellLevato91}
{Morrell}, N. \& {Levato}, H. 1991, \apjs, 75, 965

\bibitem[{{Muench} {et~al.}(2008){Muench}, {Getman}, {Hillenbrand}, \&
  {Preibisch}}]{Muenchetal08}
{Muench}, A., {Getman}, K., {Hillenbrand}, L., \& {Preibisch}, T. 2008, in
  Handbook of Star Forming Regions, Vol. I, ed. B.~{Reipurth} (San Francisco:
  ASP), 483

\bibitem[{Nelder \& Mead(1965)}]{NeMe65}
Nelder, J.~A. \& Mead, R. 1965, Comput. J., 7, 308

\bibitem[{{Nieva} \& {Przybilla}(2006)}]{NiPr06}
{Nieva}, M.~F. \& {Przybilla}, N. 2006, ApJ, 639, L39

\bibitem[{{Nieva} \& {Przybilla}(2007)}]{NiPr07}
{Nieva}, M.~F. \& {Przybilla}, N. 2007, \aap, 467, 295

\bibitem[{{Nieva} \& {Przybilla}(2008)}]{NiPr08}
{Nieva}, M.~F. \& {Przybilla}, N. 2008, \aap, 481, 199

\bibitem[{{Nieva} \& {Przybilla}(2012)}]{NiPr12}
{Nieva}, M.~F. \& {Przybilla}, N. 2012, \aap, 539, A143

\bibitem[{{Nieva} \& {Sim{\'o}n-D{\'\i}az}(2011)}]{NiSi11}
{Nieva}, M.~F. \& {Sim{\'o}n-D{\'\i}az}, S. 2011, \aap, 532, A2

\bibitem[{{Pavlovski} \& {Hensberge}(2005)}]{PaHe05}
{Pavlovski}, K. \& {Hensberge}, H. 2005, \aap, 439, 309

\bibitem[{{Pavlovski} \& {Southworth}(2009)}]{PaSo09}
{Pavlovski}, K. \& {Southworth}, J. 2009, \mnras, 394, 1519

\bibitem[{{Pavlovski} {et~al.}(2018){Pavlovski}, {Southworth}, \&
  {Tamajo}}]{Pavlovskietal18}
{Pavlovski}, K., {Southworth}, J., \& {Tamajo}, E. 2018, \mnras, 481, 3129

\bibitem[{{Pavlovski} {et~al.}(2023){Pavlovski}, {Southworth}, {Tkachenko},
  {Van Reeth}, \& {Tamajo}}]{Pavlovskietal23}
{Pavlovski}, K., {Southworth}, J., {Tkachenko}, A., {Van Reeth}, T., \&
  {Tamajo}, E. 2023, \aap, 671, A139

\bibitem[{{Preibisch} {et~al.}(1999){Preibisch}, {Balega}, {Hofmann},
  {Weigelt}, \& {Zinnecker}}]{Preibischetal99}
{Preibisch}, T., {Balega}, Y., {Hofmann}, K.-H., {Weigelt}, G., \& {Zinnecker},
  H. 1999, \na, 4, 531

\bibitem[{{Przybilla}(2005)}]{przybilla05}
{Przybilla}, N. 2005, \aap, 443, 293

\bibitem[{{Przybilla} \& {Butler}(2001)}]{PrBu01}
{Przybilla}, N. \& {Butler}, K. 2001, \aap, 379, 955

\bibitem[{{Przybilla} \& {Butler}(2004)}]{PrBu04}
{Przybilla}, N. \& {Butler}, K. 2004, \apj, 609, 1181

\bibitem[{{Przybilla} {et~al.}(2001{\natexlab{a}}){Przybilla}, {Butler},
  {Becker}, \& {Kudritzki}}]{Przybillaetal01a}
{Przybilla}, N., {Butler}, K., {Becker}, S.~R., \& {Kudritzki}, R.~P.
  2001{\natexlab{a}}, \aap, 369, 1009

\bibitem[{{Przybilla} {et~al.}(2000){Przybilla}, {Butler}, {Becker},
  {Kudritzki}, \& {Venn}}]{Przybillaetal00}
{Przybilla}, N., {Butler}, K., {Becker}, S.~R., {Kudritzki}, R.~P., \& {Venn},
  K.~A. 2000, \aap, 359, 1085

\bibitem[{{Przybilla} {et~al.}(2001{\natexlab{b}}){Przybilla}, {Butler}, \&
  {Kudritzki}}]{Przybillaetal01b}
{Przybilla}, N., {Butler}, K., \& {Kudritzki}, R.~P. 2001{\natexlab{b}}, \aap,
  379, 936

\bibitem[{{Przybilla} {et~al.}(2010){Przybilla}, {Firnstein}, {Nieva},
  {Meynet}, \& {Maeder}}]{Przybillaetal10}
{Przybilla}, N., {Firnstein}, M., {Nieva}, M.~F., {Meynet}, G., \& {Maeder}, A.
  2010, \aap, 517, A38

\bibitem[{{Przybilla} {et~al.}(2021){Przybilla}, {Fossati}, \&
  {Jeffery}}]{Przybillaetal21}
{Przybilla}, N., {Fossati}, L., \& {Jeffery}, C.~S. 2021, \aap, 654, A119

\bibitem[{{Przybilla} {et~al.}(2013){Przybilla}, {Nieva}, {Irrgang}, \&
  {Butler}}]{Przybillaetal13}
{Przybilla}, N., {Nieva}, M.~F., {Irrgang}, A., \& {Butler}, K. 2013, EAS
  Publ.~Ser., 63, 13

\bibitem[{{Raucq} {et~al.}(2018){Raucq}, {Rauw}, {Mahy}, \&
  {Sim{\'o}n-D{\'\i}az}}]{Raucqetal18}
{Raucq}, F., {Rauw}, G., {Mahy}, L., \& {Sim{\'o}n-D{\'\i}az}, S. 2018, \aap,
  614, A60

\bibitem[{{Sana} {et~al.}(2012){Sana}, {de Mink}, {de Koter}, {Langer},
  {Evans}, {Gieles}, {Gosset}, {Izzard}, {Le Bouquin}, \&
  {Schneider}}]{Sanaetal12}
{Sana}, H., {de Mink}, S.~E., {de Koter}, A., {et~al.} 2012, Science, 337, 444

\bibitem[{{Seaton}(1962)}]{Seaton62}
{Seaton}, M.~J. 1962, in Atomic and Molecular Processes, ed. D.~R. {Bates}
  (Academic Press, New York), 375

\bibitem[{{Seaton} {et~al.}(1994){Seaton}, {Yan}, {Mihalas}, \&
  {Pradhan}}]{Seatonetal94}
{Seaton}, M.~J., {Yan}, Y., {Mihalas}, D., \& {Pradhan}, A.~K. 1994, \mnras,
  266, 805

\bibitem[{{Shultz} {et~al.}(2019){Shultz}, {Le Bouquin}, {Rivinius}, {Wade},
  {Kochukhov}, {Alecian}, {Petit}, {Pfuhl}, {Karl}, {Gao}, {Grellmann}, {Lin},
  {Garcia}, {Lacour}, {MiMeS Collaboration}, \& {BinaMIcS
  Collaboration}}]{Shultzetal19}
{Shultz}, M., {Le Bouquin}, J.~B., {Rivinius}, T., {et~al.} 2019, \mnras, 482,
  3950

\bibitem[{{Simon} \& {Sturm}(1994)}]{SiSt94}
{Simon}, K.~P. \& {Sturm}, E. 1994, \aap, 281, 286

\bibitem[{{Sim{\'o}n-D{\'\i}az} {et~al.}(2011){Sim{\'o}n-D{\'\i}az},
  {Garc{\'\i}a-Rojas}, {Esteban}, {Stasi{\'n}ska}, {L{\'o}pez-S{\'a}nchez}, \&
  {Morisset}}]{Simon-Diazetal11}
{Sim{\'o}n-D{\'\i}az}, S., {Garc{\'\i}a-Rojas}, J., {Esteban}, C., {et~al.}
  2011, \aap, 530, A57

\bibitem[{{Sim{\'o}n-D{\'\i}az} {et~al.}(2006){Sim{\'o}n-D{\'\i}az}, {Herrero},
  {Esteban}, \& {Najarro}}]{Simon-Diazetal06}
{Sim{\'o}n-D{\'\i}az}, S., {Herrero}, A., {Esteban}, C., \& {Najarro}, F. 2006,
  \aap, 448, 351

\bibitem[{{Sim{\'o}n-D{\'\i}az} \& {Stasi{\'n}ska}(2011)}]{SiSt11}
{Sim{\'o}n-D{\'\i}az}, S. \& {Stasi{\'n}ska}, G. 2011, \aap, 526, A48

\bibitem[{{Smith}(1996)}]{Smith96}
{Smith}, K.~C. 1996, \apss, 237, 77

\bibitem[{{Tkachenko} {et~al.}(2016){Tkachenko}, {Matthews}, {Aerts},
  {Pavlovski}, {P{\'a}pics}, {Zwintz}, {Cameron}, {Walker}, {Kuschnig},
  {Degroote}, {Debosscher}, {Moravveji}, {Kolbas}, {Guenther}, {Moffat},
  {Rowe}, {Rucinski}, {Sasselov}, \& {Weiss}}]{Tkatchenkoetal16}
{Tkachenko}, A., {Matthews}, J.~M., {Aerts}, C., {et~al.} 2016, \mnras, 458,
  1964

\bibitem[{{Torres} {et~al.}(2010){Torres}, {Andersen}, \&
  {Gim{\'e}nez}}]{Torresetal10}
{Torres}, G., {Andersen}, J., \& {Gim{\'e}nez}, A. 2010, \aapr, 18, 67

\bibitem[{{Tully} {et~al.}(1990){Tully}, {Seaton}, \&
  {Berrington}}]{Tullyetal90}
{Tully}, J.~A., {Seaton}, M.~J., \& {Berrington}, K.~A. 1990, J. Phys. B, 23,
  3811

\bibitem[{{Valencic} {et~al.}(2004){Valencic}, {Clayton}, \&
  {Gordon}}]{Valencicetal04}
{Valencic}, L.~A., {Clayton}, G.~C., \& {Gordon}, K.~D. 2004, \apj, 616, 912

\bibitem[{{van Regemorter}(1962)}]{vanRegemorter62}
{van Regemorter}, H. 1962, \apj, 136, 906

\bibitem[{{Vrancken} {et~al.}(1996){Vrancken}, {Butler}, \&
  {Becker}}]{Vranckenetal96}
{Vrancken}, M., {Butler}, K., \& {Becker}, S.~R. 1996, \aap, 311, 661

\bibitem[{{We{\ss}mayer} {et~al.}(2022){We{\ss}mayer}, {Przybilla}, \&
  {Butler}}]{Wessmayeretal22}
{We{\ss}mayer}, D., {Przybilla}, N., \& {Butler}, K. 2022, \aap, 668, A92

\bibitem[{{We{\ss}mayer} {et~al.}(2024){We{\ss}mayer}, {Urbaneja}, {Butler}, \&
  {Przybilla}}]{Wessmayeretal24}
{We{\ss}mayer}, D., {Urbaneja}, M.~A., {Butler}, K., \& {Przybilla}, N. 2024,
  \aap, 687, L7

\bibitem[{{Wiese} {et~al.}(1996){Wiese}, {Fuhr}, \& {Deters}}]{Wieseetal96}
{Wiese}, W.~L., {Fuhr}, J.~R., \& {Deters}, T.~M. 1996, J. Phys. \& Chem. Ref.
  Data., Monograph 7 (AIP Press, Melville, NY)

\end{thebibliography}

\onecolumn
\begin{appendix}
\section{Spectral fit of HD~37061}\label{appendix:A}

\begin{figure*}[ht]
    \centering
    \includegraphics[width=0.995\hsize]{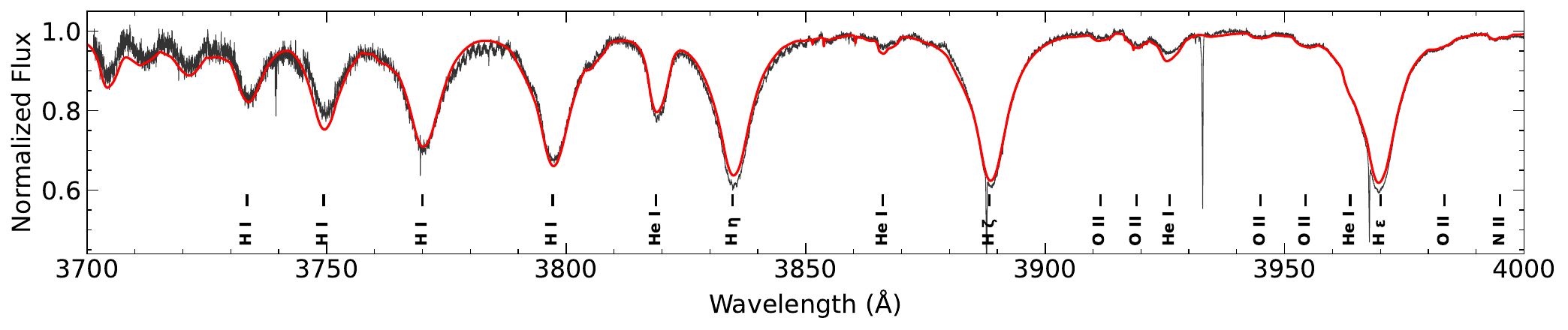}
    \includegraphics[width=0.995\hsize]{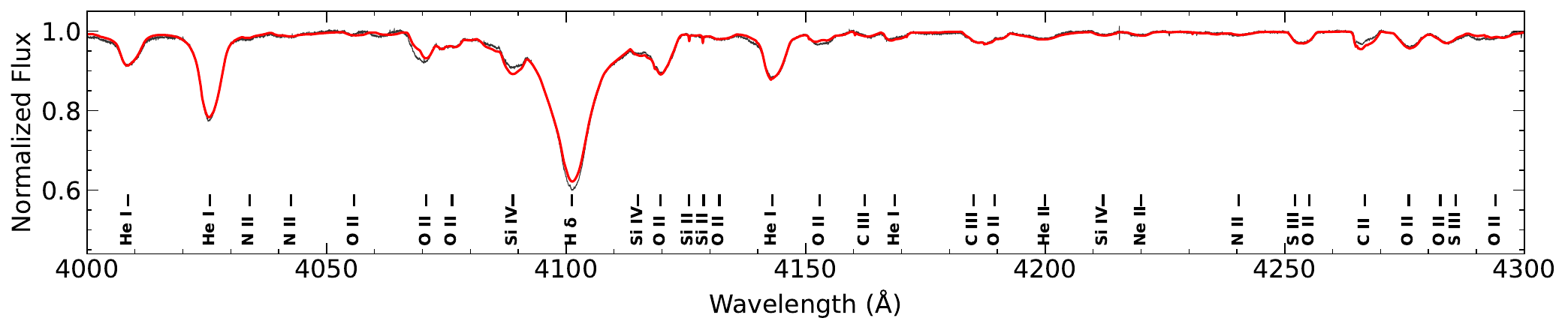}
    \includegraphics[width=0.995\hsize]{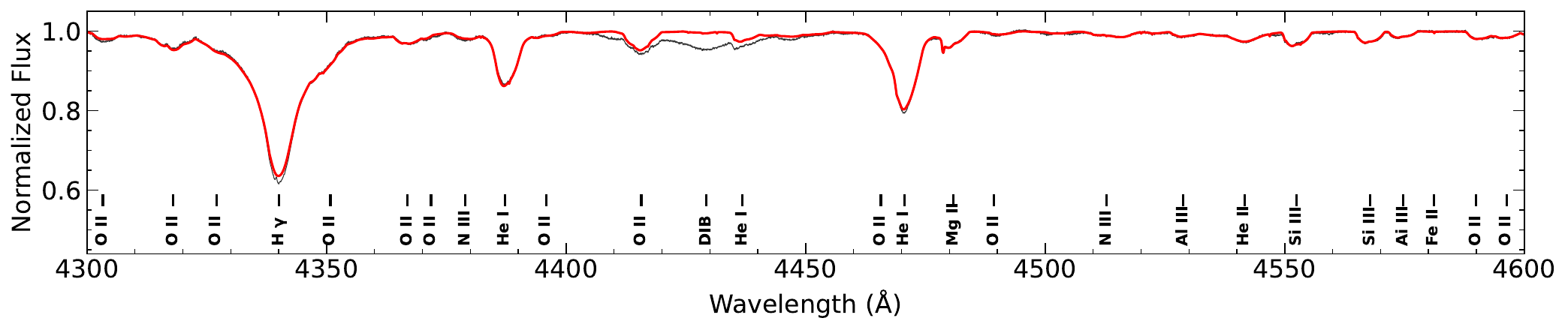}
    \includegraphics[width=0.995\hsize]{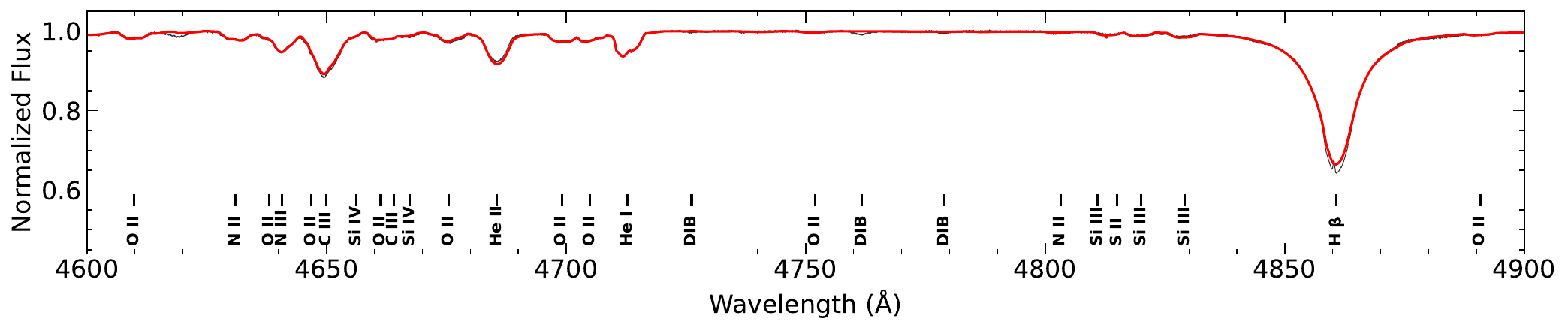}
    \includegraphics[width=0.995\hsize]{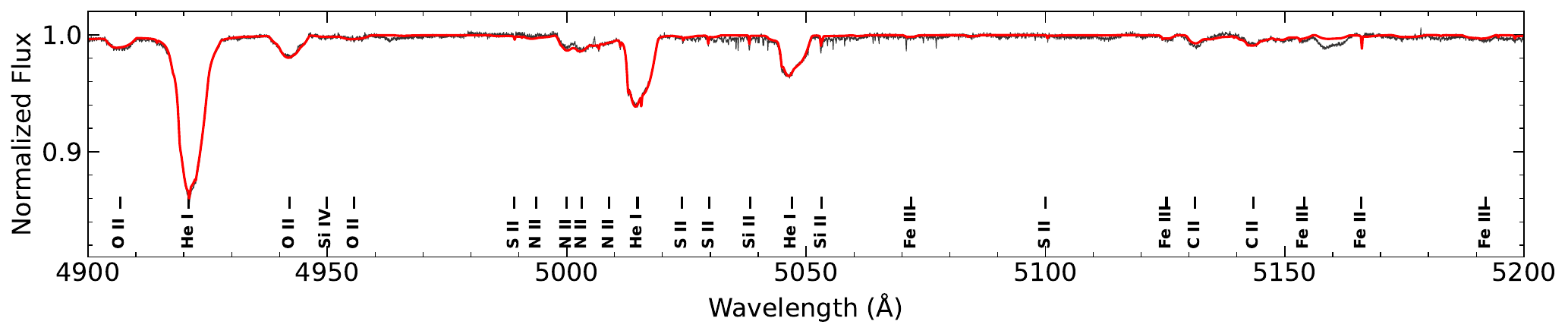}
    \caption{Comparison between the observed spectrum (black) and the global best fitting model (red). The spectrum is shifted into the rest frame of HD\,37061\,Aa. Many of the stronger diagnostic lines and diffuse interstellar bands (DIBs) are identified.}
    \label{fig:full_fit}
\end{figure*}

\addtocounter{figure}{-1}
\begin{figure*}[ht]
    \centering
    \includegraphics[width=0.995\hsize]{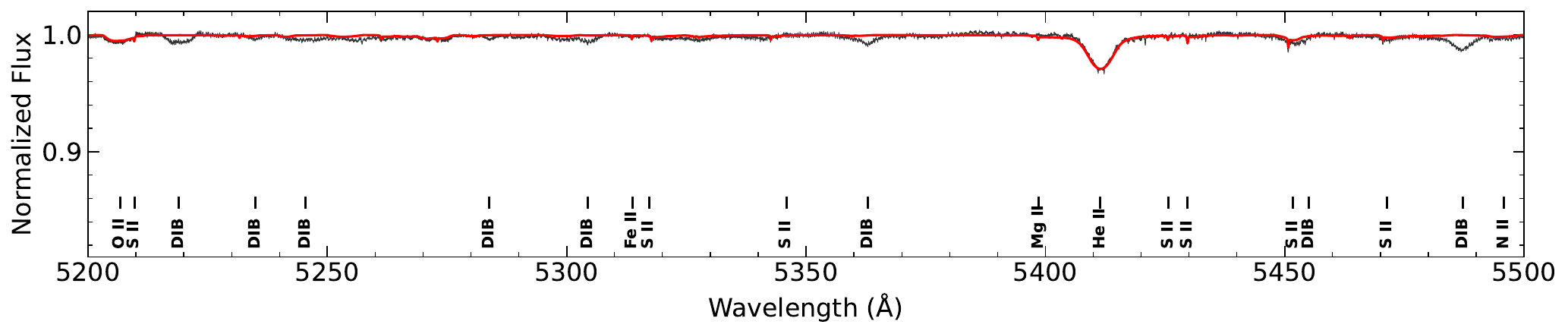}
    \includegraphics[width=0.995\hsize]{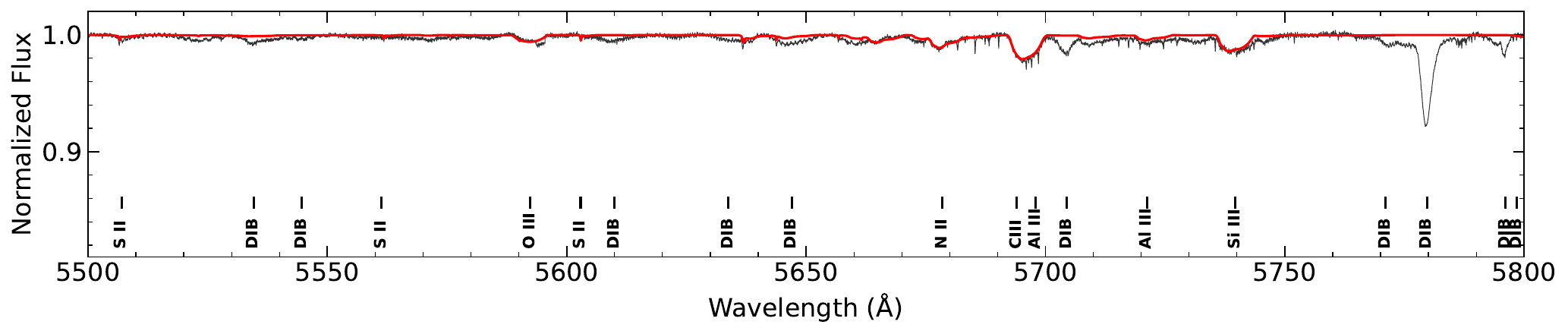}
    \includegraphics[width=0.995\hsize]{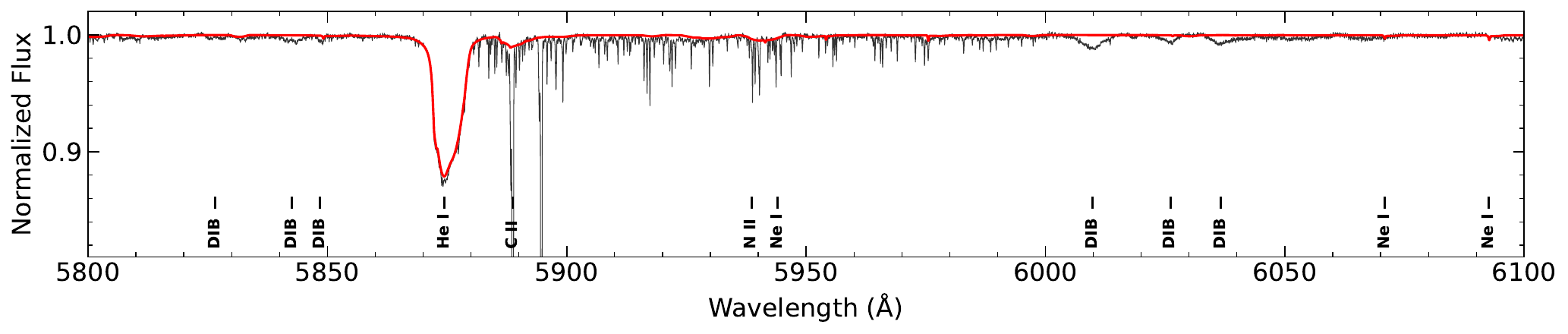}
    \includegraphics[width=0.995\hsize]{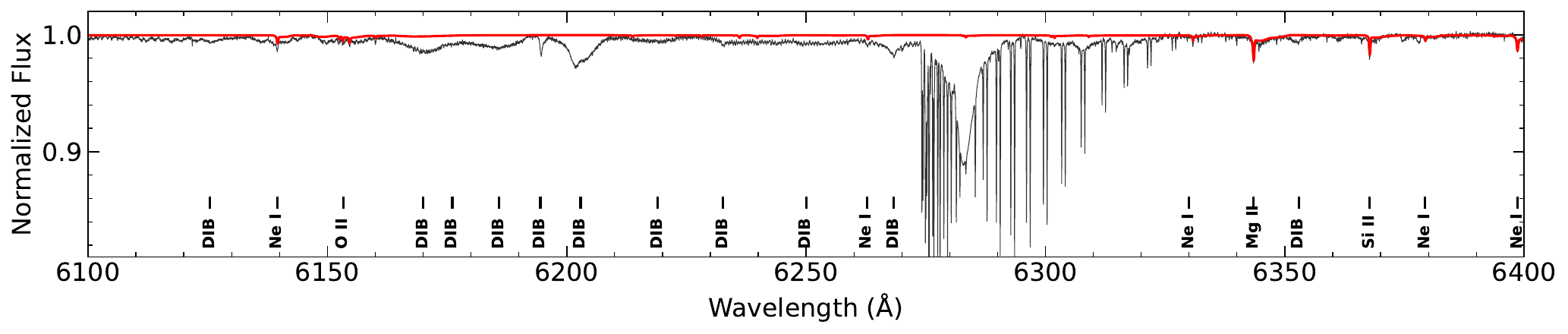}
    \includegraphics[width=0.995\hsize]{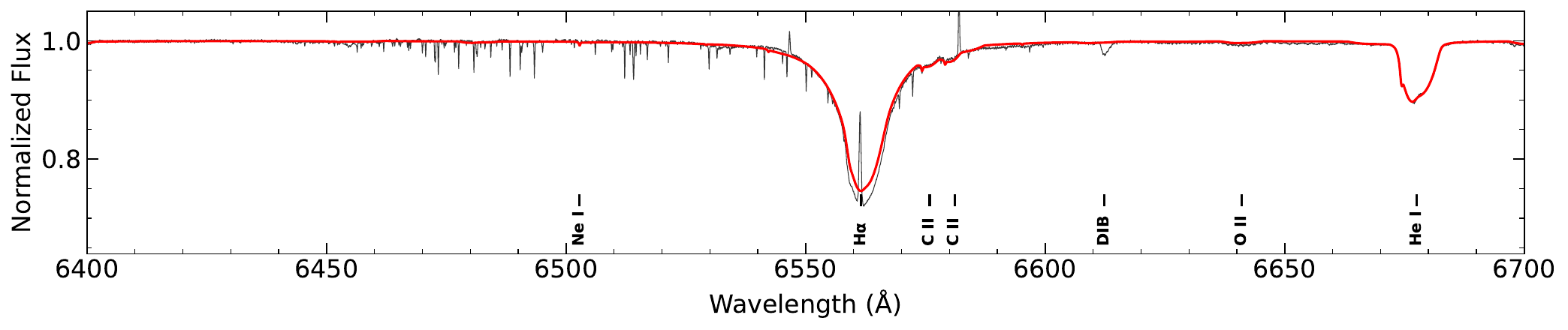}
    \includegraphics[width=0.995\hsize]{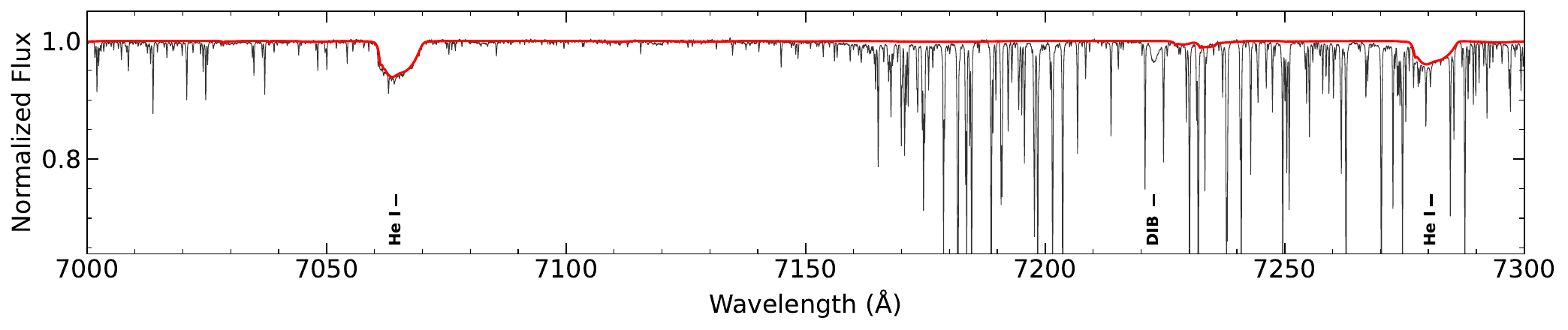}
    \caption{Continued.}
\end{figure*}

\end{appendix}

\end{document}